\documentclass[preprintnumbers,amsmath,amssymb,floatfix,nofootinbib,aps]{revtex4}
\pdfoutput=1
\usepackage{amsmath,amsfonts,bbm,euscript,hyperref}
\usepackage{graphicx}
\usepackage{dcolumn}
\usepackage{bm}
\usepackage{epsfig}
\epsfclipon

\begin{document}

\preprint{UMN-TH-3112/12}

\title{Curvature Perturbations from a Massive Vector Curvaton}

\author{Ryo Namba\footnote{Electronic address: namba@physics.umn.edu}}
\affiliation{School of Physics and Astronomy,University of Minnesota, Minneapolis, 55455 (USA)}

\date{\today}

\begin{abstract} 
We study a ghost-free model of massive vector curvaton proposed in the literature, where the quick decrease of the vector background expectation value is avoided by a suitable choice of kinetic and mass functions. The curvaton perturbations of this model have been so far computed assuming that these functions are external classical quantities, and it was found that some special time evolution of these functions leads to scale invariant and statistically isotropic perturbations of the vector curvaton. However,  external functions should be understood as originating from the expectation value of some additional field. Since these functions need to present a non-trivial evolution during inflation, the field cannot be trivially integrated out, and, in particular, its perturbations need to be included in the computation. We do so in a minimal implementation of the mechanism, where the additional field is identified with the inflaton. We show that, except for a narrow window of model parameters, the interaction with this field generally causes the curvature perturbations to violate statistical isotropy beyond the observational limit. 
\end{abstract}

\maketitle

\begin{widetext}
\tableofcontents\vspace{5mm}
\end{widetext}

\section{Introduction} 
\label{sec:introduction}

In the past three decades, primordial inflation has become a dominant paradigm for the very early universe. It provides a simple, yet compelling mechanism to resolve the conceptual problems that the standard Big Bang cosmology confronts, namely the horizon, flatness, and monopole problems \cite{Guth:1980zm, Linde:1981mu, Albrecht:1982wi}. On the other hand, the observations of large-scale structure (LSS) and cosmic microwave background (CMB) exhibit small fluctuations of order $10^{-5}$, which are considered of primordial origin. The same theories that realize inflation can also predict such fluctuations with the observed features (near scale invariance, nearly Gaussian statistics, greater power in scalar modes than in tensor), which makes inflation a particularly favorable theory (see e.g. \cite{Lyth:1998xn} for review). However, many realizations of inflation give degenerate predictions compatible with these observations, and therefore there is a large parameter space yet to explore for the underlying particle physics in the very early universe. Considerable amounts of study have been done to discriminate among otherwise degenerate models. One discriminator is non-Gaussianity, deviation from the Gaussian statistics, in the CMB. While non-Gaussianity is predicted to be undetectable in the simplest models of inflation \cite{Acquaviva:2002ud, Maldacena:2002vr, Seery:2005wm, Seery:2008qj}, a number of extended models have been studied which predict detectable non-Gaussian signature \cite{Chen:2006nt, Barnaby:2007yb, Karciauskas:2008bc, Meerburg:2009ys, Langlois:2009jp, Barnaby:2010ke, Barnaby:2010vf}. The current bounds on non-Gaussianity are still loose \cite{Komatsu:2010fb}, but future missions, such as Planck satellite, are expected to improve measurement precisions to  unprecedented level \cite{Liguori:2010hx}.

Observable non-Gaussianity, and other interesting signatures, can be obtained in the curvaton mechanism \cite{Lyth:2001nq}, where the field responsible for the primordial perturbations is not the inflaton. While for simplicity it is most often assumed that the curvaton is a scalar field, models of vector curvatons have also been proposed \cite{Dimopoulos:2009am, Dimopoulos:2009vu, vector-curvaton}. Vector fields break the background isotropy through their vacuum expectation value (vev), and therefore their spectrum of perturbations may break statistical isotropy. Interestingly, a $30\%$ level  violation of statistical isotropy  has been detected at $9 \sigma$  in the WMAP data  \cite{Hanson:2009gu, Groeneboom:2009cb}. However, the direction of the asymmetry nearly coincides with the ecliptic one, which strengthens the case for a systematic origin of the effect. Fortunately, this can be checked by Planck, that can be sensitive to anisotropies at the few percent level \cite{Pullen:2007tu}. The observed anisotropy, and the expected improvement from Planck, have motivated the study of cosmological models that can produce this effect, including those of vector curvaton.

In the standard scenarios with an expansion by a (approximate) positive cosmological constant, it has been shown that the models of almost all Bianchi types evolve rapidly toward the de Sitter solution and any anisotropy present initially will be washed away (sometimes called cosmic no-hair conjecture) \cite{Wald:1983ky}. To obtain an observable anisotropy, the premises in \cite{Wald:1983ky} have to be violated. One way to accomplish this is to introduce a vector field with a non-vanishing  vev during inflation. 

However, in the case of a familiar U(1) gauge field with a minimal kinetic term $- F^2 / 4$ and no potential term, the field is conformally coupled to gravity. Thus its energy density quickly decays away and cannot source any appreciable anisotropy for sufficiently long time. By breaking the conformal invariance, the vector field can provide non-vanishing prolonged shear pressure, which supports a prolonged stage of anisotropic expansion and leads to violation of statistical isotropy. Several implementations of this idea have been suggested \cite{Ackerman:2007nb, Ford:1989me}; however, these models break the U(1) symmetry and inevitably introduce a longitudinal polarization which, for these models, happens to be a ghost  \cite{Himmetoglu:2008zp, Himmetoglu:2008hx, Himmetoglu:2009qi}. To avoid this instability, a model was suggested in \cite{Watanabe:2009ct}, and further studied in \cite{Dulaney:2010sq,Gumrukcuoglu:2010yc, Watanabe:2010fh}, where the U(1) gauge invariance is preserved and therefore no dangerous longitudinal mode is present. In this work, the fast decrease of the gauge field vev is avoided by introducing a scalar function that multiplies the vector kinetic term, ${\cal L} \supset - f \left( \varphi \right) F^2 / 4$, where $\varphi$ is the inflaton field. \footnote{The kinetic term used in  \cite{Watanabe:2009ct} is analogous to the one originally employed by Ratra \cite{Ratra:1991bn}. The magnetogenesis application of \cite{Ratra:1991bn} suffers of a strong coupling problem
\cite{Demozzi:2009fu,Barnaby:2012tk}; however it was shown in  \cite{Barnaby:2012tk} that the model \cite{Ratra:1991bn} (without need of identifying the vector field with the electromagnetic one) produces non-Gauassianity of the primordial perturbations of nearly local type.}

A similar idea to that of  \cite{Watanabe:2009ct} was used in \cite{Dimopoulos:2009am, Dimopoulos:2009vu}, where the vector field plays the role of the curvaton, and where two time dependent functions were considered
\footnote{
More recently \cite{Dimopoulos:2012av}, this model was extended to  includes the effects from a parity violating term,  proportional  $F \tilde{F}$, where $\tilde{F}$ is the dual of the field-strength tensor. The non-Gaussian signatures due to particle production through the interaction $\varphi F \tilde{F}$, where $\varphi$ is a pseudo-scalar inflaton, are studied extensively in \cite{Barnaby:2010vf}.
}
\begin{equation}
{\cal L} =  - \frac{f \left( t \right)}{4} F_{\mu \nu} F^{\mu \nu} - \frac{1}{2} m^2 \left( t \right) A_{\mu} A^{\mu}  \label{lag-original}
\end{equation}
The kinetic coupling is typical of moduli or dilaton-like fields in string theory and supergravity frameworks, and the mass can be induced by a Higgs-like mechanism. Since the vector field is massive, the additional degree of freedom (longitudinal mode) is present, but, for $f,m^2 >0$ the model has no ghosts. It is shown in \cite{Dimopoulos:2009am, Dimopoulos:2009vu} that the curvature perturbations generated in this model attain the scale-invariant statistically-isotropic power spectrum, if the varying functions have a specific time dependence, i.e. $f \propto a^{-4}$ and $m \propto a$, where $a$ is the scale factor, and if the vector mass is initially light and is heavy by the end of inflation (the physical vector mass $m/\sqrt{f}$ grows considerably during inflation).
 
In a field theory, an external time dependence should be understood as the vev of a field. Given that the functions $f$ and $m$ needs to present a nontrivial evolution during inflation, this field cannot be trivially integrated out, and, in particular, its perturbations cannot be disregarded. The field acts as ``clock''; therefore, the most minimal choice is to identify this field with the inflaton field. Ref. \cite{Dimopoulos:2010xq} already provided some specific examples where $f$ and $m$ are functions of the inflaton, and showed that,  for a generic inflaton potential that satisfies the slow-roll conditions, the attractor solutions of the background evolution lead to small anisotropy in the expansion and to the required time dependence for the background values of $f$ and $m$. However, ref. \cite{Dimopoulos:2010xq} does not compute cosmological perturbations in this set-up, but refers to the results of 
 \cite{Dimopoulos:2009am, Dimopoulos:2009vu}, in which the functions $f$ and $m$ are only external classical quantities.

The non-minimal coupling through the kinetic and mass terms modulated by inflaton inevitably induces an interaction between the scalar and the vector perturbations. It is natural to think that this interaction modifies the evolution of perturbations in a non-trivial way. As we will show later in this paper, this interaction, which is present already in the linearized level, can actually be directionally biased due to the background anisotropy. Moreover, since the system of perturbations is now a coupled one, consistent quantization has to be done in the matrix form, the formulation first developed in \cite{Nilles:2001fg} and summarized in Section \ref{sec:perturbations}. In this paper, we treat the full coupled system consistently, and show that the scalar-vector interaction produces direction-dependent effects in the power spectrum of curvature perturbations, in the framework of the vector curvaton scenario. We find that the near scale invariance of the spectrum is a generic feature of the model.  The (near) statistical isotropy can also be achieved; however, it requires a specific choice of the coupling constant entering in the functions $f$ and $m$, and appropriate initial conditions. This is in contrast with the approximated computation of \cite{Dimopoulos:2009am, Dimopoulos:2009vu}, in which the statistical isotropy appeared to be a generic result, independent of the functional form of $f$ and $m$. 

We stress that, despite introducing an inflaton, we still want to work under the assumption of  \cite{Dimopoulos:2009am,Dimopoulos:2009vu} that this is a model of vector curvaton, so that the vector field should be  responsible for the cosmological perturbations. This can for instance be achieved by assuming that, after inflation, the inflaton quickly decays into relativistic fields, and that the energy density of the decay products become completely negligible with respect to that of the curvaton (or its decay products). We do no follow the details of reheating here, but we simply compute the vector field perturbations until they freeze out, and we assume that they are the source of cosmological perturbations, according to the curvaton mechanism hypothesis. As also done in  \cite{Dimopoulos:2009am,Dimopoulos:2009vu}, we still disregard (for technical reasons, as the computation is very involved) the metric perturbations in the analysis. These are the same working assumptions of the realizations of this mechanism suggested in \cite{Dimopoulos:2010xq}. In addition to  \cite{Dimopoulos:2010xq}, we however consistently include the interaction with the inflaton perturbations induced by the two functions $f$ and $m$  that characterize this model. We show that this drastically changes the curvaton power spectrum in this model.

The paper is organized as follows. In Section \ref{sec:background}, we analyze the background dynamics including the vev of the scalar and vector fields and the anisotropic expansion of the universe. The background attractor of this model is derived. In Section \ref{sec:perturbations}, we discuss the evolution of the perturbations. We focus on the 2D scalar modes, which are the ones that contribute to the energy density and thus to the curvature perturbations. We quantize the system consistently for a coupled system and derive the equations of motion to evolve the system of perturbations in the matrix form. In Section \ref{sec:observables}, we numerically compute the power spectrum of the vector density fluctuations and relate it to that of curvature perturbations. Results are shown for some different values of parameters. In Section \ref{sec:conclusions}, we discuss the results and conclude. Throughout the paper, natural units are used, $\hbar = c = 1$, and the Einstein notation is assumed for repeated indices.

\section{Background Dynamics}
\label{sec:background}

We consider a model with a scalar inflaton $\varphi$ and a massive vector field $A_\mu$ whose mass and kinetic functions vary in time. This model is a realization of the mechanism of \cite{Dimopoulos:2009am,Dimopoulos:2009vu}, where cosmological perturbations were computed  under the assumption that the vector field is a curvaton, and using the Lagrangian (\ref{lag-original}).  The computation of  \cite{Dimopoulos:2009am,Dimopoulos:2009vu} can be considered as an approximated computation of the perturbations, performed under the following non-trivial simplifying 
assumptions: (i) the background dynamics is in pure de Sitter, (ii) the background anisotropy induced from the vev of the vector field is negligible, (iii) the vector mass and kinetic functions are classical external functions of time, with no  perturbations. The goal of this work is to improve over  the results of  \cite{Dimopoulos:2009am,Dimopoulos:2009vu} by removing these assumptions.

\subsection{Model and Setup}

The model we consider is described by the action
\begin{equation}
S = \int d^4 x \, \sqrt{-g} \left[ \frac{M_p^2}{2} R - \frac{1}{2} \partial_{\mu} \varphi \, \partial^{\mu} \varphi - V \left( \varphi \right) - \frac{f \left( \varphi \right)}{4} F_{\mu \nu} F^{\mu \nu} - \frac{1}{2} m^2 \left( \varphi \right) A_{\mu} A^{\mu} \right] \label{action}
\end{equation}
where $F_{\mu\nu} = \partial_\mu A_\nu - \partial_\nu A_\mu$ and $M_p$ is the reduced Planck mass. The universe expansion during inflation is driven by $\varphi$ with a sufficiently flat potential $V \left( \varphi \right)$, allowing slow-roll inflation (which we assume throughout this paper). Here the vector mass and kinetic terms are function of $\varphi$ rather than being  external functions of time. In \cite{Dimopoulos:2009am, Dimopoulos:2009vu}, it is claimed that the scale invariance and statistical isotropy of the power spectrum of primordial curvature perturbations can be achieved in the vector curvaton scenario, if $f$ and $m$ have time dependence $f \propto a^{-4}$ and $m \propto a$, where $a$ is the overall scale factor. With an appropriate choice of $f \left( \varphi \right)$ and $m \left( \varphi \right)$, these time dependences can be dynamically achieved (as shown in \cite{Dimopoulos:2010xq}, and as we review below  for our particular model).

We can, without loss of generality, fix the background coordinate system such that the vector vev points in the $x$ direction, i.e. $\left< A_\mu \right> = \left( 0, A \left( t \right), 0, 0 \right)$. A background metric consistent with this choice is a Bianchi type I metric with the residual isotropy in the $yz$ plane, given by
\begin{equation}
ds^2 = -dt^2 + a^2\left( t \right) \, dx^2 +  b^2\left( t \right) \left( dy^2 + dz^2 \right), \label{metric}
\end{equation}
where we parametrize the two scale factors by $a\left( t \right) = {\rm e}^{\alpha(t) - 2 \sigma(t)}$ and $b\left( t \right) = {\rm e}^{\alpha(t) + \sigma(t)}$. Hence $\alpha \left( t \right)$ is the number of e-folds averaged over all directions, and $\sigma \left( t \right)$ measures the difference in e-folds between the two directions. Note that $\dot{\alpha}$ corresponds to the ``overall'' Hubble parameter, and $\dot{\sigma}$ corresponds to the shear.

There are four dynamical degrees of freedom in the background system of this model, $\phi \left( t \right)$, $A \left( t \right)$, $\alpha \left( t \right)$ and $\sigma \left( t \right)$, where $\phi \left( t \right)$ is the homogeneous vev of $\varphi$. The corresponding equations of motion are, respectively,
\begin{eqnarray}
&& \ddot{\phi} + 3 \, \dot{\alpha} \, \dot{\phi} + V' = f \, {\rm e}^{-2 \alpha + 4 \sigma} \left( \frac{f'}{2f} \dot{A}^2 - \frac{m \, m'}{f} A^2 \right) \label{eqphi} \\
&& \ddot{A} + \left( \dot{\alpha} + 4 \, \dot{\sigma} +\frac{f'}{f} \dot{\phi} \right) \dot{A} + \frac{m^2}{f} A = 0 \label{eqA} \\
&& 2 \, \ddot{\alpha} + 3 \, \dot{\alpha}^2 + 3 \, \dot{\sigma}^2 = \frac{1}{M_p^2} \left( -\frac{1}{2} \dot{\phi}^2 + V \right) + \frac{f}{6 M_p^2} {\rm e}^{-2 \, \alpha + 4 \, \sigma} \left( - \dot{A}^2 + \frac{m^2}{f} A^2 \right) \label{eqalpha} \\
&& \ddot{\sigma} + 3 \, \dot{\alpha} \, \dot{\sigma} = \frac{f}{3 M_p^2} \, {\rm e}^{-2 \alpha + 4 \sigma} \left( \dot{A}^2 - \frac{m^2}{f} A^2 \right) \label{eqsigma}
\end{eqnarray}
with the Einstein constraint equation
\begin{equation}
3 \, \dot{\alpha}^2 - 3 \, \dot{\sigma}^2 = \frac{1}{M_p^2} \left[ \left( \frac{1}{2} \, \dot{\phi}^2 + V \right) + \frac{f}{2} \, {\rm e}^{-2 \alpha + 4 \sigma} \left( \dot{A}^2 + \frac{m^2}{f} A^2 \right) \right] .  \label{constraint}
\end{equation}
Here, and throughout the paper, dot denotes the derivative with respect to $t$, and prime with respect to $\varphi$. Notice that Eq. (\ref{eqsigma}) implies that the evolution of $\sigma$ is directly supported only by the pressure of vector field. Also notice that the system is invariant under the unphysical rescalings $\alpha \rightarrow \alpha + \alpha_0$, $\sigma \rightarrow \sigma + \sigma_0$, $A \rightarrow A \, {\rm e}^{\alpha_0 - 2 \sigma_0}$, where $\alpha_0$ and $\sigma_0$ are arbitrary constants, and thus the normalization of $\alpha$ and $\sigma$ does not affect any physical quantity. As a consistency check, if we set $A = 0$ and $\dot{\sigma} = 0$, we would recover the standard single-scalar-field inflation. It is also useful to define the background energy densities for the scalar and vector fields, given by
\begin{eqnarray}
\rho_{\phi} & = & \frac{1}{2} \, \dot{\phi}^2 + V \label{rho_phi} \\
\rho_A & = & \frac{f}{2} \, {\rm e}^{-2 \, \alpha + 4 \, \sigma} \left( \dot{A}^2 + \frac{m^2}{f} A^2 \right). \label{rho_A}
\end{eqnarray}
The inflaton energy density is the standard one, and in the slow-roll regime, $V$ ($\gg \frac{1}{2} \dot{\phi}^2$) drives the quasi exponential expansion. Due to the non-zero vector mass, the vector energy density is also split into the kinetic and potential parts. Interestingly, in the pure de Sitter ($\dot{\sigma} = 0$ and $\dot{\alpha}$ is constant), $\rho_A$ is {\emph{exactly}} constant during inflation, if $f \propto a^{-4}$ and $m \propto a$. As a consequence, in the quasi de Sitter background with small anisotropy, which is the case we consider, $\rho_A$ stays nearly constant during inflation until the desired time dependence of $f$ and $m$ starts to be violated near the end of inflation. Also, it is worth noting that $A$ is not a physical quantity; it always appears in the combination $\sqrt{f} {\rm e}^{- \alpha + 2 \, \sigma} A$, which is the physical one, and the physical mass of the vector field is 
\begin{equation}
M \equiv \frac{m}{\sqrt{f}} .
\end{equation}
In the following, we occasionally use $M$ instead of $m$, when the meaning is more transparent. We should note that $M$ scales as $\propto a^3$ during inflation and that we are interested in the parameter space where $M \ll \dot{\alpha}$ at early stages of inflation and $M \gg \dot{\alpha}$ by the end of inflation, which attains the scale invariant and statistically isotropic power spectrum in the curvaton scenario for the case of isotropic de Sitter background and unperturbed $f,m$ \cite{Dimopoulos:2009am,Dimopoulos:2009vu}.

The desired time dependence of the mass and kinetic functions, $f \propto a^{-4}$ and $m \propto a$, have to be achieved dynamically. For a concrete realization of this, we consider the simplest chaotic potential of inflaton
\begin{equation}
V \left( \varphi \right) = \frac{1}{2} m_\varphi^2 \varphi^2 . \label{pot_V}
\end{equation}
With this form of potential, in order to obtain the desired time dependence, $f$ and $m$ take the form
\begin{equation}
f \left( \varphi \right) = {\rm e}^{\frac{c \, \varphi^2}{M_{p}^2}}, \quad m \left( \varphi \right) = m_0 \, {\rm e}^{- \frac{c \, \varphi^2}{4 \, M_{p}^2}}
\label{def_fncs}
\end{equation}
where $c$ and $m_0$ are positive constants. Notice that after inflation, $\phi \left( t \right)$ starts oscillating and approaches  $0$; accordingly, $f$ approaches to unity, and $m$ to a constant value $m_0$. We will now derive the attractor of the background system and show that it leads to $f \propto a^{-4}$ and $m \propto a$ during inflation.

\subsection{Attractor Solutions}

We are interested in finding the attractor in the slow-roll inflationary regime. There are four dynamical background degrees of freedom in this system, but none of them has an exact analytic solution. However, in the pure de Sitter with negligible anisotropy, (\ref{eqA}) has an analytic solution  for $A \left( t \right)$ in the case $f \propto a^{-4}$ and $m \propto a$. Motivated by this fact, we parametrize $A \left( t \right)$ and $\dot{A} \left( t \right)$ in terms of $C \left( t \right)$ and $D \left( t \right)$ in the following way:
\begin{eqnarray}
A \left( t \right) & \equiv & C \left( t \right) \, \cos \int_{t_{\rm{in}}}^t dt' M \left[ \phi \left( t' \right) \right]+ D \left( t \right) \, \sin \int_{t_{\rm{in}}}^t dt' M \left[ \phi \left( t' \right) \right] \label{Adec} \\
\dot{A} \left( t \right) & \equiv & M \left[ \phi \left( t \right) \right] \left[ - C \left( t \right) \, \sin \int_{t_{\rm{in}}}^t dt' M \left[ \phi \left( t' \right) \right] + D \left( t \right) \, \cos \int_{t_{\rm{in}}}^t dt' M \left[ \phi \left( t' \right) \right] \right] \label{Addec}
\end{eqnarray}
where $t_{\rm{in}}$ is some early time during inflation. Such parametrization can always be done. In  pure de Sitter, $C$ and $D$ are integration constants; here, they are promoted to be time dependent. In this parametrization, we can solve two first-order differential equations for $C$ and $D$, instead of one second-order equation for $A$. The former is in fact numerically favorable, since $C$ and $D$ vary little during inflation and do not have a large hierarchy between themselves. Also, we assume the initial equipartition for the vector energy density: namely, the vector kinetic energy is equal to the vector potential at initial stages of inflation (this is assumed in \cite{Dimopoulos:2009am}, and we verified that it is indeed a necessary condition for the statistical isotropy in the power spectrum in the case of unperturbed $f$ and $m$). Since $M \ll \dot{\alpha}$ at early times, we have $A \simeq C$ and $\dot{A} \simeq M D$, and thus the energy equipartition implies 
\begin{equation}
\left| C \left( t_{\rm{in}} \right) \right| = \left| D \left( t_{\rm{in}} \right) \right|.
\end{equation}
To investigate the background attractor solution, we assume $\left| C \right| \sim \left| D \right|$ and that $C$ and $D$ are constants (the precise solution of $C$ and $D$ is instead used in the cumputation of the perturbations).%
\footnote{In fact, $D$ changes some, depending on the value of $c$ and the value of $\dot{\sigma}$ (which in turn depends on the ratio between the vector kinetic and potential energies), while $C$ stays really constant until near the end of inflation.}

\begin{figure}
\centering
\includegraphics[width=0.55\textwidth]{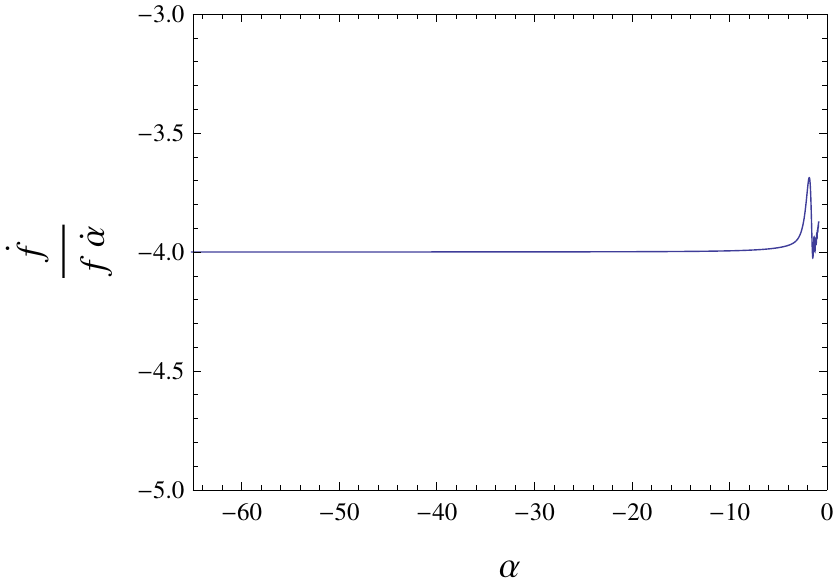}
\caption{Time dependence of $f$. The $x$-axis is the overall e-folds $\alpha$, and $y$-axis is the quantity $\dot{f} / f \dot{\alpha}$, which would be equal to $-4$ if $f \propto a^{-4}$ exactly. This desired time dependence is well retrieved until the vector field becomes heavy ($\alpha \simeq -2.2$). At $M \approx \dot{\alpha}$, the time dependence is slightly violated; before it comes back to $-4$, inflation ends ($\alpha =0$), and $f$ approaches $1$. The parameters here are taken to be $c = 1.5$ and $m_0 = 1000 \, m_\varphi$. By construction, $m$ obeys $\propto a$ when $f \propto a^{-4}$.}
\label{fig:f_timedep}
\end{figure}

We focus on the background equations for $\phi$, $\sigma$ and the constraint, namely (\ref{eqphi}), (\ref{eqsigma}) and (\ref{constraint}), respectively. Here we consider the early time regime, namely the vector mass is light $M \ll \dot{\alpha}$. From (\ref{eqphi}), ignoring $\ddot{\phi}$ and $\sigma$ terms (slow roll and small anisotropy), we have
\begin{equation}
3 \, \dot{\alpha}\, \dot{\phi} \simeq \left[ - m_\varphi^2 + \frac{c \, m_0^2 \left( C^2 + 2 D^2 \right)}{2 \, M_p^2} \, {\rm e}^{-2 \alpha - \frac{c \, \phi^2}{2 \, M_p^2}} \right] \phi.
\end{equation}
Now we use $\alpha$ as the time variable with $\frac{d}{dt} = \dot{\alpha} \frac{d}{d \alpha}$. Since the right-hand side of (\ref{constraint}) is dominated by the term $V$ (which will be verified), we have $\dot{\alpha}^2 \simeq \frac{1}{3 \, M_p^2} V$. Defining $\tilde{\phi}^2 \equiv \phi^2 + \frac{4 \, M_p^2}{c} \alpha$, we write the above equation in terms of $\tilde{\phi}$,
\begin{equation}
\tilde{\phi} \, \frac{d \tilde{\phi}}{d \alpha} \simeq - 2 \, M_p^2 \frac{c - 1}{c} + c \, \frac{m_0^2}{m_\varphi^2} \left( C^2 + 2 D^2 \right) {\rm e}^{- \frac{c \, \tilde{\phi}^2}{2 \, M_p^2}} . \label{att_phitil}
\end{equation}
The homogeneous solution for this equation implies that the time variation of $\tilde{\phi}$ is small, so we can neglect the left-hand side of (\ref{att_phitil}). Then we find
\begin{equation}
{\rm e}^{- \frac{c \, \tilde{\phi}^2}{2 \, M_p^2}} = {\rm e}^{-2 \alpha - \frac{c \, \phi^2}{2 \, M_p^2}} \simeq \frac{c - 1}{c^2} \, \frac{2 \, M_p^2}{C^2 + 2 D^2} \, \frac{m_\varphi^2}{m_0^2} \label{attractor1}
\end{equation}
Since the right-hand side of (\ref{attractor1}) is constant, we obtain $\alpha \simeq - \frac{c \, \phi^2}{4 \, M_p^2}$, up to some constant. This means that the chosen forms of $f$ and $m$ indeed obey the desired time dependence, $f \propto a^{-4}$ and $m \propto a$, in this attractor. This is verified numerically in FIG. \ref{fig:f_timedep}. Combining this with $\dot{\alpha}^2 \simeq \frac{1}{6 \, M_p^2} m_\varphi^2 \phi^2$ gives
\begin{equation}
3 \, \dot{\alpha} \, \dot{\phi} \simeq - \frac{m_\varphi^2 \phi}{c}, \quad {\rm{or}} \quad \dot{\phi} \simeq - \frac{m_\varphi M_p}{c} \sqrt{\frac{2}{3}} . \label{att_phid}
\end{equation}
This coincides with the attractor found in \cite{Watanabe:2009ct}. It is worth noting that for this attractor solution to be valid, we need to have $c > 1$, as can easily be seen in (\ref{attractor1}). From here on, we assume this range of values of $c$.

\begin{figure}
\centering
\includegraphics[width=0.6\textwidth]{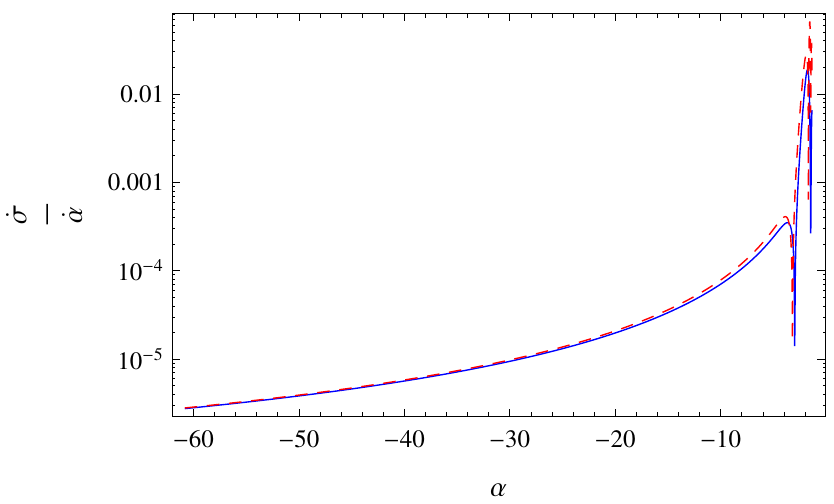}
\caption{Evolution of the anisotropy, $\vert \dot{\sigma} \vert / \dot{\alpha}$. The numerical solution (blue solid curve) is compared to the attractor solution (\ref{att_sigd}) (red dashed curve), where the numerical solutions are used for $r_A$ and $\phi$. Here the overall e-folding ($\alpha$) is normalized to $0$ when inflation ends. The vector field becomes heavy ($M = \dot{\alpha}$) at $\alpha \simeq -2.2$. As discussed in the main text, the attractor solution matches well with the numerical until $M \approx \dot{\alpha}$; after this moment, the anisotropy decays quickly, but (\ref{att_sigd}) does not replicate that. Here the parameters are taken to be $c = 1.5$, $m_0 = 1000 \, m_\varphi$ and $r_A^{\rm{initial}} = 1$. For other values of parameters, the attractor is valid as long as $c$ is not too close to $1$.}
\label{fig:att_sigma}
\end{figure}

To consider the anisotropy, we ignore the $\ddot{\sigma}$ term in (\ref{eqsigma}) and take the phase in the parametrization of $A$ and $\dot{A}$ to be zero. Since we assume initial equipartition for the vector energy density, it is convenient to define 
\begin{equation}
r_A \equiv \frac{\rho_A^V}{\rho_A^{\rm{kin}}}
\end{equation}
i.e. the ratio of the vector potential to the vector kinetic energy density. We are interested in the case where this ratio stays close to $1$. At initial times, this ratio reduces to $r_A^{\rm{initial}} \simeq \frac{C^2}{D^2}$. Combining these with (\ref{attractor1}) and $\dot{\alpha}^2 \simeq \frac{1}{6 \, M_p^2} m_\varphi^2 \phi^2$, we find the attractor lead the anisotropy to
\begin{equation}
\frac{\dot{\sigma}}{\dot{\alpha}} \simeq - \frac{4}{3} \, \frac{c - 1}{c^2} \, \frac{r_A - 1}{r_A + 2} \, \frac{M_p^2}{\phi^2} . \label{att_sigd}
\end{equation}
Notice that $\dot{\sigma}$ vanishes if $r_A = 1$. This is clear both intuitively and from (\ref{eqsigma}). If $\rho_A^V = \rho_A^{\rm{kin}}$, the right-hand side of (\ref{eqsigma}) becomes $0$, and then $\dot{\sigma}$ decays away to $0$ quickly. The only matter content in this model that causes the shear pressure is the vector field. When the vector kinetic and potential terms are equal, the pressure vanishes, like non-relativistic dust, and does not enhance anisotropy. In the actual evolution, $r_A$ does not stay exactly equal to $1$, so anisotropy is produced to a small amount, {\emph{until the vector field becomes heavy}}. Once the vector mass exceeds the background expansion rate ($M > \dot{\alpha}$), the pressure vanishes and so does the anisotropy. FIG. \ref{fig:att_sigma} shows that (\ref{att_sigd}) is a very good approximation, until the vector field becomes heavy ($M \approx \dot{\alpha}$).

We can also determine from the attractor the ratio between the energy densities of the scalar and the vector. Assuming slow roll and small anisotropy, we find
\begin{equation}
\frac{\rho_A}{\rho_{\phi}} \simeq 2 \, \frac{c - 1}{c^2} \, \frac{r_A + 1}{r_A + 2} \, \frac{M_p^2}{\phi^2} .
\label{att_energy}
\end{equation}
We note that in the limit of $r_A \rightarrow 0$ (massless), this recovers the attractor solution found in \cite{Watanabe:2009ct, Gumrukcuoglu:2010yc} and one of the attractor solutions in \cite{Dimopoulos:2010xq}; on the other hand, in the limit of $r_A \gg 1$, it recovers another attractor in \cite{Dimopoulos:2010xq}. We stress that our early-time attractor expressions (\ref{att_phid}), (\ref{att_sigd}) and (\ref{att_energy}) are valid for arbitrary values of $r_A$.

Our starting assumptions are now verified: anisotropy is small ($\dot{\sigma} \ll \dot{\alpha}$), and the vector energy density is subdominant ($\rho_A \ll \rho_\varphi$), in the slow-roll regime with the potential $V$ of the form (\ref{pot_V}). The vector kinetic and mass functions of the form (\ref{def_fncs}) are dynamically led to their desired time dependence by the attractor. The existence of the attractor requires $c > 1$, and we only consider this range of $c$. Also, we assume the initial equipartition for the vector density, which translates to $\left| C \left( t_{\rm{in}} \right) \right| = \left| D \left( t_{\rm{in}} \right) \right|$ as an initial condition. As long as anisotropy is small with quasi de Sitter expansion and $f$ and $m$ obey the time dependence, $C$ and $D$ are nearly constant, and the equipartition is maintained. In fact, if $c$ is too close to $1$, the attractor is no longer a good approximation; then the time dependence of $f$ and $m$ starts differing from the desired one, and the initial equipartition deviates as the system evolves in time.

\section{Perturbations}
\label{sec:perturbations}

We now consider the perturbations of the model (\ref{action}). To avoid an extreme computational complication, we neglect metric perturbations in this paper. While a further study including them would be interesting, we will show that the inclusion of background anisotropy and the perturbations of $\varphi$ drastically changes the power spectrum of curvature perturbations, as compared to the case of isotropic, pure de Sitter background with no $\delta \varphi$, studied in \cite{Dimopoulos:2009am}. Here we study the effects in the linearized level of perturbation theory.

\subsection{Basics}

First note that since the vector mass is put by hand, there is no U(1) gauge freedom (so we never call it ``gauge'' field). Consequently the vector perturbations contain three independent degrees of freedom, two transverse and one longitudinal, in terms of momentum directions. In some of massive vector models, the longitudinal modes may suffer from ghost instabilities \cite{Himmetoglu:2008zp, Himmetoglu:2008hx, Himmetoglu:2009qi}; however, this particular model we consider here is ghost-free \cite{Dimopoulos:2009vu}.

Since we are neglecting the metric perturbations, the metric still takes the form (\ref{metric}). We define the field perturbations as the field values with their background subtracted,
\begin{equation}
\delta \varphi \left( t , \vec{x} \right) \equiv \varphi \left( t , \vec{x} \right) - \left< \varphi \right> \;, \quad \delta A_\mu \left( t , \vec{x} \right) \equiv A_\mu \left( t , \vec{x} \right) - \left< A_\mu \right>
\end{equation}
and their Fourier transformations through
\begin{equation}
\delta \varphi \left( t, \vec{x} \right) = \int{\frac{d^3k}{\left(2 \pi \right)^{3/2}} \, {\rm e}^{i \vec{k} \cdot \vec{x}} \, \delta \varphi \left( t, \vec{k} \right)} \;, \quad \delta A_\mu \left( t, \vec{x} \right) = \int{\frac{d^3k}{\left(2 \pi \right)^{3/2}} \, {\rm e}^{i \vec{k} \cdot \vec{x}} \, \delta A_\mu \left( t, \vec{k} \right)}
\end{equation}
as usual. To simplify the computation and to make geometrical interpretation transparent, we orient the coordinate system such that the $z$-component of the comoving momentum vanishes \cite{Gumrukcuoglu:2007bx},
\begin{equation}
\vec{k} = \left( k_L, k_T, 0 \right) \label{k_orient}
\end{equation}
where $L$ and $T$ denote ``longitudinal'' and ``transverse'' in terms of the background vector vev. This choice can be made without loss of generality, due to the residual 2D symmetry in $yz$ plane after orienting the vector vev along the $x$-axis (as done in Section \ref{sec:background}). In this coordinate system, it can easily be seen that the 2nd-order action is split into two decoupled sectors, 2D scalar ($\left\{ \delta \varphi, \delta A_0, \delta A_x, \delta A_y \right\}$) and 2D vector ($\left\{ \delta A_z \right\}$). The 2D vector mode $\delta A_z$ does not contribute to the curvature perturbations, which are intrinsically a scalar quantity (this is more explicitly shown in Section \ref{sec:observables}), and thus we disregard this sector in the following discussions. We denote the physical momentum $\vec{p} = \left( p_L , p_T , 0 \right)$, with its components and norm
\begin{equation}
p_{L} = \frac{k_{L}}{a} \;, \quad p_{T} = \frac{k_{T}}{b} \;, \quad p = \sqrt{p_L^2 + p_T^2} .
\end{equation}
(Recall $a = {\rm e}^{\alpha - 2 \sigma}$ and $b = {\rm e}^{\alpha + \sigma}$.)

Among the 2D scalar modes, $\delta A_0$ is non-dynamical and can be integrated away. We see that $\delta A_0$ is related to other 2D scalar modes by
\begin{equation}
\delta A_0 = \frac{-i}{p^2 + M^2} \left( \frac{p_L}{a} \, \delta \dot{A}_x + \frac{p_T}{b} \, \delta \dot{A}_y + \frac{f'}{f} \, \frac{p_L}{a} \, \dot{A} \, \delta \varphi  \right) . \label{nondyn}
\end{equation}
We now have three independent modes in 2D scalar and one in 2D vector, as expected. One of the two vector transverse mode is the 2D vector mode $\delta A_z$; the other transverse and the longitudinal modes are linear combinations of $\delta A_x$ and $\delta A_y$. The 2D scalar sector of 2nd-order action is now
\begin{eqnarray}
S_{\rm{2dS}}^{(2)} & = & \frac{1}{2} \int{dt \, d^3 k} \, {\rm e}^{3 \alpha} \Bigg\{ \left| \delta \dot{\varphi} \right|^2 + \frac{f}{a^2} \, \frac{p_T^2 + M^2}{p^2 + M^2} \left| \delta \dot{A}_x \right|^2 + \frac{f}{b^2} \, \frac{p_L^2 + M^2}{p^2 + M^2} \left| \delta \dot{A}_y \right|^2  - \frac{f}{a b} \, \frac{p_L \, p_T}{p^2 + M^2} \left( \delta \dot{A}_x^{\dagger} \, \delta \dot{A}_y + \delta \dot{A}_y^{\dagger} \, \delta \dot{A}_x \right) \nonumber \\
	&&  + \frac{f'}{a^2} \frac{p_T^2 + M^2}{p^2 + M^2} \dot{A} \left( \delta \varphi^{\dagger} \, \delta \dot{A}_x + \delta \dot{A}_x^{\dagger} \delta \varphi  \right) - \frac{f'}{a b} \frac{p_L p_T}{p^2 + M^2} \dot{A} \left( \delta \varphi^{\dagger} \, \delta \dot{A}_y + \delta \dot{A}_y^{\dagger} \delta \varphi  \right)  \nonumber \\
	&& - \left[ p^2 + V'' + \frac{f}{a^2} \, \dot{A}^2 \left( \frac{f'^2}{f^2} \frac{p_L^2}{p^2 + M^2} - \frac{f''}{2 \, f} \right) + \frac{f}{a^2} \, M^2 A^2 \left( \frac{M'^2}{M^2} + \frac{f''}{2 \, f} + \frac{2 \, f' \, M'}{f \, M} + \frac{M''}{M} \right) \right] \left| \delta \varphi \right|^2 \nonumber \\
	&& - \frac{f}{a^2} \left( p_T^2 + M^2 \right) \left| \delta A_x \right|^2 - \frac{f}{b^2} \left( p_L^2 + M^2 \right) \left| \delta A_y \right|^2 \nonumber \\
	&& - \frac{f}{a^2} \, M^2 A \left( \frac{f'}{f} + 2 \frac{M'}{M} \right) \left( \delta \varphi^{\dagger} \, \delta A_x + \delta A_x^{\dagger} \, \delta \varphi \right) + \frac{f}{a b} \, p_L p_T \left( \delta A_x^{\dagger} \, \delta A_y + \delta A_y^{\dagger} \, \delta A_x \right) \Bigg\} .
\label{action_2dS}
\end{eqnarray}
We need to solve this system to find the power spectrum of curvature perturbations. However, it is not possible to solve such a complicated system exactly; moreover, we look for the effects from non-zero $\dot{\sigma}$ and from the coupling between $\delta A_\mu$ and $\delta \varphi$, and thus approximated methods would be quite delicate. Thus we employ numerical computation here to obtain the final results. To this end, we need to appropriately quantize the system and find the initial conditions. This is a coupled system, and so the quantization and the evolution have to be done in the matrix form, as described in detail in \cite{Gumrukcuoglu:2010yc, Nilles:2001fg} and summarized in the following subsections.

\subsection{Quantization}

We first rotate the system such that the kinetic mixing term present in the first line of (\ref{action_2dS}) vanish. This can be done through
\begin{equation}
\delta_i = {\cal R}_{ij} \hat{Y}_j 
\end{equation}
where $\delta_i = \left( \delta \varphi , \delta A_x , \delta A_y \right)_i$ and a rotation matrix 
\begin{equation}
{\cal R}_{ij} = {\rm e}^{-3 \alpha /2} \left( \begin{array}{ccc}
1 & 0 & 0 \\
0 & \frac{a}{\sqrt{p_{T}^2 + M^2}} \, c_{T} & \frac{a}{\sqrt{p_{T}^2 + M^2}} \, c_{L} \\
0 & -\frac{b}{\sqrt{p_{L}^2 + M^2}} \, c_{T} & \frac{b}{\sqrt{p_{L}^2 + M^2}} \, c_{L} \\
\end{array}\right)_{ij}
\label{rotation}
\end{equation}
with arbitrary time dependent functions $c_T$ and $c_L$. By choosing them as
\begin{equation}
c_T^2 = \frac{p^2 + M^2}{2 f \left[ 1 + p_L \, p_T / \sqrt{\left( p_L^2 + M^2 \right) \left( p_T^2 + M^2 \right)} \right]} \;, \quad c_L^2 = \frac{p^2 + M^2}{2 f \left[ 1 - p_L \, p_T / \sqrt{\left( p_L^2 + M^2 \right) \left( p_T^2 + M^2 \right)} \right]}
\label{cTL_quant}
\end{equation}
then the kinetic matrix becomes the identity. Subtracting appropriate total derivatives, the action (\ref{action_2dS}) takes the form
\begin{equation}
S_{\rm{2dS}}^{(2)} = \frac{1}{2} \int{dt \, d^3 k} \left[ \dot{\hat{Y}}^\dagger \, \dot{\hat{Y}} + \dot{\hat{Y}}^\dagger \, \hat{X} \, \hat{Y} - \hat{Y}^\dagger \, \hat{X} \, \dot{\hat{Y}} - \hat{Y}^\dagger \, \hat{\Omega}^2 \, \hat{Y} \right] \label{action_quant}
\end{equation}
where all the components of the matrices $\hat{X}$ and $\hat{\Omega}^2$ are real. These matrices have further properties $\hat{X}^T = - \hat{X}$ and $\hat{\Omega}^{2 \, T} = \hat{\Omega}^2$, which is always possible by subtracting total derivatives; moreover, $\hat{X}$ and $\hat{\Omega}^2$ are invariant under the parity $\vec{k} \rightarrow - \vec{k}$. Also note that all the field components entering in the action satisfies $\hat{Y}_i^\dagger \left( - \vec{k} \right) =  \hat{Y}_i \left( \vec{k} \right)$ (so that $\hat{Y}_i$ are real in the coordinate space). The action (\ref{action_quant}) is formally of the same type as in \cite{Gumrukcuoglu:2010yc}, so we follow the procedure outlined there.

To remove the $\hat{X}$ terms, define
\begin{equation}
\psi = R \, \hat{Y}
\end{equation}
where $R$ is a real, orthogonal matrix, so that $\hat{Y}^\dagger \, \hat{Y} = \psi^\dagger \psi$, and satisfies the relation $\dot{R} = R \, \hat{X}$. The action now becomes
\begin{equation}
S = \frac{1}{2}\int{dt \, d^3k} \left[ \dot{\psi}^\dagger \dot{\psi} - \psi^\dagger \tilde{\Omega}^2 \psi \right] \label{action2}
\end{equation}
where $\tilde{\Omega}^2 = R \left( \hat{\Omega}^2 - \hat{X}^2 \right) R^T$. Note that we can always choose $R$ to have the property $R \left( - \vec{k} \right) = R \left( \vec{k} \right)$ since $\hat{X}$ satisfies the same property. Hence, $\tilde{\Omega}^2 \left( - \vec{k} \right) = \tilde{\Omega}^2 \left( \vec{k} \right)$, and $\psi^\dagger \left( - \vec{k} \right) = \psi \left( \vec{k} \right)$. Thanks to this property, it is immediately realized that $\psi$ is an array of real fields in the coordinate space and that the conjugate momentum is defined as $\pi_i \equiv \dot{\psi}_i$. The Hamiltonian is then $H = \frac{1}{2} \int{d^3k} \left[ \pi^\dagger \pi + \psi^\dagger \tilde{\Omega}^2 \psi \right]$. To diagonalize the Hamiltonian, we further define 
\begin{equation}
\psi = C \, \hat{\psi} \;, \quad \pi = C \, \hat{\pi}
\end{equation}
where $C$ is an orthogonal matrix ($C^T \, C = 1$) that diagonalize $\tilde{\Omega}^2$, i.e. 
\begin{equation}
C^T \, \tilde{\Omega}^2 \, C = {\rm{diag}} \left( \omega_1^2 , \omega_2^2, \omega_3^2 \right) \equiv \omega^2 
\end{equation}
giving $H = \frac{1}{2} \int{d^3 k} \left[ \hat{\pi}^\dagger \hat{\pi} + \hat{\psi}^\dagger \omega^2 \hat{\psi} \right]$. Note that, since $\tilde{\Omega}^2$ is a Hermitian matrix and $\tilde{\Omega}^2 \left( - \vec{k} \right) = \tilde{\Omega}^2 \left( \vec{k} \right)$, $\omega_i^2$ are all real and $\omega^2 \left( - \vec{k} \right) = \omega^2 \left( \vec{k} \right)$. Since $C$ is unchanged under the parity $\vec{k} \rightarrow - \vec{k}$, we then define
\begin{eqnarray}
\psi \left( \vec{k} \right) & = & C_{ij} \left( \vec{k} \right) \left[ h_{jl} \left( \vec{k} \right) P_{lm} \, a_m \left( \vec{k} \right) + h_{jl}^{*}\left( - \vec{k} \right) P_{lm}^* \, a_m^\dagger \left( - \vec{k} \right) \right] \nonumber \\
\pi \left( \vec{k} \right) & = & C_{ij} \left( \vec{k} \right) \left[ \tilde{h}_{jl} \left( \vec{k} \right) P_{lm} \, \left( \vec{k} \right) a_m \left( \vec{k} \right) + \tilde{h}_{jl}^{*} \left( - \vec{k} \right) P_{lm}^* \, a_m^\dagger \left( - \vec{k} \right) \right] \label{psipi_h}
\end{eqnarray}
where $a_i \left( \vec{k} \right)$ and $a_i^\dagger \left( \vec{k} \right)$ are arrays of annihilation and creation operators, respectively, satisfying the commutation relation $\left[ a_i \left( \vec{k}_1 \right) , a_j^\dagger \left( \vec{k}_2 \right) \right] = \delta_{ij} \delta^{(3)} \left( \vec{k}_1 - \vec{k}_2 \right)$. The matrix $P_{ij}$ is a constant Hermitian matrix; this phase freedom is present in quantizing a coupled system, similarly to arbitrary phases appearing in a free-field theory, and this phase does not affect any physical result (in fact, any consistent quantization should be constructed allowing such freedom). In the coordinate space, $\psi$ and $\pi$ are real, and to have $\psi$ and $\pi$ satisfy the equal-time commutation relation, we require
\begin{equation}
\left[ h \left( \vec{k} \right) \tilde{h}^\dagger \left( \vec{k} \right) - h^* \left( - \vec{k} \right) \tilde{h}^T \left( - \vec{k} \right) \right]_{ij} = i \, \delta_{ij} . \label{ETCR}
\end{equation}
We decompose $h$ and $\tilde{h}$ in a similar manner to Bogolyubov transformation: 
\begin{equation}
h = \frac{1}{\sqrt{2 \, \omega}} \left( \alpha + \beta \right) \;, \quad \tilde{h} = \frac{- i \, \omega}{\sqrt{2 \, \omega}} \left( \alpha - \beta \right) .
\end{equation}
Here $h$, $\tilde{h}$, $\omega$, $\alpha$ and $\beta$ are all matrices, but since $\omega^{2}$
is a diagonal matrix, $1 / \sqrt{2 \, \omega}$ and $\omega / \sqrt{2 \, \omega}$ are well defined matrices. With these ``Bogolyubov coefficients'' $\alpha$ and $\beta$, the condition (\ref{ETCR}) is translated to
\begin{equation}
\left[ \alpha \left( \vec{k} \right) \alpha^\dagger \left( \vec{k} \right) - \beta^* \left( - \vec{k} \right) \beta^T \left( - \vec{k} \right) \right]_{ij} = \delta_{ij} \; , \quad \left[ \alpha \left( \vec{k} \right) \beta^\dagger \left( \vec{k} \right) - \beta^* \left( - \vec{k} \right) \alpha^T \left( - \vec{k} \right) \right]_{ij} = 0 . \label{ETCR2}
\end{equation}
This is not the unique choice to satisfy (\ref{ETCR}), but it is a sufficient condition and the following derivation is consistent with this choice. The Hamiltonian now becomes, after
normal-ordering,
\begin{equation}
: \! H \! : = \int{d^3 k} \, \omega_i \, b_i^\dagger \left( \vec{k} \right) \, b_i \left( \vec{k} \right)
\end{equation}
where $b_{i}$ and $b_{i}^{\dagger}$ are defined as
\begin{equation}
\left[ \begin{array}{c}
	b \left( \vec{k} \right) \\
	b^\dagger \left( - \vec{k} \right)
	\end{array}%
\right]%
= \left[ \begin{array}{cc}
	\alpha \left( \vec{k} \right) & \beta^* \left( - \vec{k} \right) \\
	\beta \left( \vec{k} \right) & \alpha^* \left( - \vec{k} \right)
	\end{array}%
\right]%
\left[ \begin{array}{c}
	a \left( \vec{k} \right) \\
	a^\dagger \left( - \vec{k} \right)
\end{array} \right] .
\end{equation}
Using (\ref{ETCR2}), it can be shown $b$ and $b^{\dagger}$ satisfies $\left[ b_i \left( \vec{k}_1 \right) , b_j^\dagger \left( \vec{k}_2 \right) \right] = \delta_{ij} \, \delta^{(3)} \left( \vec{k}_1 - \vec{k}_2 \right)$. The Hamiltonian is thus fully diagonalized and the system is consistently quantized.

We are now ready to find the equations of motion for the system in terms of $\left\{ h , \tilde{h} \right\}$ or of $\left\{ \alpha , \beta \right\}$. As we have $\dot{\psi}_i \left( \vec{k} \right) = \pi_i \left( \vec{k} \right)$ and
$\dot{\pi}_i \left( \vec{k} \right) = - \tilde{\Omega}^2 \left( \vec{k} \right) \psi \left( \vec{k} \right)$ (from (\ref{action2})), the equations of motion for $h$ and $\tilde{h}$ are 
\begin{equation}
\dot{h} = \tilde{h} - \Gamma \, h \;, \quad \dot{\tilde{h}} = - \Gamma \, \tilde{h} - \omega^2 h \label{EOMh}
\end{equation}
where $\Gamma = C^T \dot{C}$. With $\alpha$ and $\beta$, these equations are expressed as
\begin{equation}
\dot{\alpha} = - i \, \omega \, \alpha + \frac{\dot{\omega}}{2 \, \omega} \, \beta - I \alpha - J \beta \;, \quad \dot{\beta} = i \, \omega \, \beta + \frac{\dot{\omega}}{2 \, \omega} \, \alpha - I \beta - J \alpha
\end{equation}
where 
\begin{equation}
I = \frac{1}{2} \left[ \sqrt{\omega} \, \Gamma \frac{1}{\sqrt{\omega}} + \frac{1}{\sqrt{\omega}} \, \Gamma \sqrt{\omega} \right] \;, \quad J = \frac{1}{2} \left[ \sqrt{\omega} \, \Gamma \frac{1}{\sqrt{\omega}} - \frac{1}{\sqrt{\omega}} \, \Gamma \sqrt{\omega} \right].
\end{equation}
At asymptotically early times, when modes are deeply inside the horizon, $\hat{\Omega}_{ij}^2 \simeq p^2 \delta_{ij}$
and $\hat{X}_{ij}\ll p$, and consequently $\omega_i \simeq p$ and $\Gamma , I , J , \frac{\dot{\omega}}{\omega} \ll \omega$.
Thus we obtain the adiabatic initial condition $\alpha_{\rm{in}} = {\rm e}^{- i \int^t dt' \omega_{\rm{in}}} \mathbbm{1}$ and $\beta_{\rm{in}} = 0$, or equivalently
\begin{eqnarray}
h_{\rm{in}} & = & \frac{1}{\sqrt{2 \, \omega_{\rm{in}}}} {\rm e}^{- i \int^t dt' \omega_{\rm{in}}} \simeq \frac{1}{\sqrt{2 \, p_{\rm{in}}}} {\rm e}^{- i \int^t dt' \omega_{\rm{in}}} \mathbbm{1} \nonumber \\
\tilde{h}_{\rm{in}} & = & \frac{-i \, \omega_{\rm{in}}}{\sqrt{2 \, \omega_{\rm{in}}}} {\rm e}^{- i \int^t dt' \omega_{\rm{in}}} \simeq - i \sqrt{\frac{p_{\rm{in}}}{2}} {\rm e}^{- i \int^t dt' \omega_{\rm{in}}} \mathbbm{1} . \label{initialh}
\end{eqnarray}
This is the solution to (\ref{EOMh}) at early times and satisfies the quantization condition (\ref{ETCR}). It corresponds to the initial adiabatic vacuum, if we choose the Bunch-Davis initial vacuum which is annihilated by the operator $a_i$ appearing in (\ref{psipi_h}). Therefore, we have formally quantized the coupled system of the 2D scalar sector of the model, and we can determine the initial conditions through (\ref{initialh}) to evolve the system.

\subsection{Early-Time Evolution}

We now need to evolve the system given by the action (\ref{action_2dS}), with the initial conditions determined by (\ref{initialh}). As is clear from the quantization procedure, this must be done in the matrix form. It would be ideal to rotate the system using the rotation matrix (\ref{rotation}) with the coefficients $c_{T , L}$ in (\ref{cTL_quant}) to make the kinetic matrix the identity, and to evolve $h_{ij}$ and $\tilde{h}_{ij}$ according to their equations of motion (\ref{EOMh}). However, we have found it computationally challenging to use (\ref{cTL_quant}). To circumvent this issue, we still rotate the system by (\ref{rotation}) but choose $c_{T,L}$ such that the kinetic matrix becomes very close to the identity {\emph{only}} at early times $M \ll p$. For this purpose, we take
\begin{equation}
c_{T} = \frac{p}{2 \sqrt{f}} \; , \quad c_{L} = \frac{p_{L} \, p_{T}}{M \sqrt{f}} \label{cTL} .
\end{equation}
The action (\ref{action_2dS}) now becomes
\footnote{Here, $X$ is not an antisymmetric matrix, i.e. $X^T \ne - X$. Although anytisymmetrization is always possible for $X$, we did not do it here, only for simpler expressions of the matrices, which are explicitly written in Appendix \ref{appA}. This does not change the evolution, since antisymmetrization is done by adding total derivatives in the action and thus does not affect the equations of motion.}
\begin{equation}
S^{(2)}_{\rm{2dS}} = \frac{1}{2} \int{d t \, d^3 k} \left[ \dot{Y}^{\dagger} \, T \, \dot{Y} + \dot{Y}^{\dagger} \, X \, Y + Y^{\dagger} \, X^{T} \, \dot{Y} - Y^{\dagger} \, \Omega^2 \, Y \right] \label{action_early} .
\end{equation}
Here, each component of $Y_i$ is related to that of $\hat{Y}_i$ in the previous subsection only by rescaling due to a different choice of the coefficients. The matrix $T$ is diagonal, and at early times, we have $T_{\rm{early}} \simeq \mathbbm{1}$ as we required; moreover, $\Omega_{\rm{early}}^{2} \simeq p^2 \, \mathbbm{1}$ and $X_{\rm{early}} \ll p$, necessary for adiabatic initial conditions. The complete expressions for the matrices $T$, $X$ and $\Omega^2$ are written in Appendix \ref{appA}. From the expression of $T_{33}$, it is manifest that $Y_3$ corresponds to the longitudinal mode, with the free choice of ${\rm{sign}} \left( p_L \, p_T \right) > 0$.

We evolve the system of $Y_i$. By varying the action (\ref{action_early}), the equations of motion are obtained as
\begin{equation}
\ddot{Y}_{i} + {\mathcal A}_{ij} \, \dot{Y}_{j} + {\mathcal B}_{ij} \, Y_{j} = 0 \label{EOM_early}
\end{equation}
where
\begin{equation}
{\mathcal A} = T^{-1} \left( \dot{T} + X - X^{T} \right) \; , \quad {\mathcal B} = T^{-1} \left( \Omega^2 + \dot{X} \right) . \label{matAB}
\end{equation}
Since $T$ is diagonal, $T^{-1}$ is trivially found. We now need to decompose $Y_i$ into arrays of creation and annihilation operators with matrix coefficients, consistent with the quantization procedure, as in (\ref{psipi_h}). That is,
\begin{equation}
Y_i \left( t, \vec{k} \right) = {\mathcal Y}_{ij} \left( t, \vec{k} \right) a_j \left( \vec{k} \right) + {\mathcal Y}_{ij}^* \left( t, - \vec{k} \right) a_j^\dagger \left( - \vec{k} \right) \label{dec_Y}
\end{equation}
where the creation and annihilation operators here are the same as those that enter in (\ref{psipi_h}). The matrix coefficients ${\mathcal Y}_{ij}$  are related to $h_{ij}$ by
\begin{equation}
{\mathcal Y}_{ij} = \left( T^{-1/2} R^T C h \right)_{il} P_{lj} \; , \quad \dot{{\mathcal Y}}_{ij} = \left[ T^{-1/2} R^T C \, \tilde{h}  + \partial_t \left( T^{-1/2} R^T C \right) h \right]_{il} P_{lj} \label{Ytoh}
\end{equation}
where the matrices $R$, $C$, and $P$ are introduced in the previous subsection. By writing in terms of ${\cal Y}$, the equations of motion (\ref{EOM_early}) consist of two parts, one proportional to $a_j \left( \vec{k} \right)$ and the other to $a_j^\dagger \left( - \vec{k} \right)$, and these two terms must vanish simultaneously. They are in fact not independent but equivalent terms, due to the fact that ${\mathcal A}$ and ${\mathcal B}$ are real and invariant under $\vec{k} \rightarrow - \vec{k}$. Therefore the equation of motion we need to evolve in the matrix form is
\begin{equation}
\ddot{{\mathcal Y}}_{ij} + {\mathcal A}_{il} \, \dot{{\mathcal Y}}_{lj} + {\mathcal B}_{il} \, {\mathcal Y}_{lj} = 0. \label{EOM_Y}
\end{equation}

To determine the initial conditions, we employ the adiabatic initial condition (\ref{initialh}) and relate it to ${\mathcal Y}$ through (\ref{Ytoh}). In doing so, we take advantage of the phase freedom by fixing the arbitrary constant Hermitian matrix $P$ to eliminate the initial phase. (This procedure is done in \cite{Watanabe:2010fh}.) Namely, we choose $P = C^T R$ at initial time; since both of $R$ and $C$ are orthogonal, and since both $h$ and $\tilde{h}$ are proportional to the identity, we can simplify (\ref{Ytoh}) to ${\mathcal Y} = T^{-1/2} \, h$ and $\dot{{\mathcal Y}} \simeq T^{-1/2} \, \tilde{h}  + n \, \dot{\alpha} \, T^{-1/2} \, h$, where $n$ is some number of ${\mathcal O} \left( 1 \right)$. Recall that $T_{\rm{early}} \simeq \mathbbm{1}$ with subdominant terms completely negligible (suppressed by $\sim M^2 / p^2$), and $\partial_t \left( T^{-1/2} R^T C \right) \sim \dot{\alpha} \,T^{-1/2} R^T C$. The second term in $\dot{{\mathcal Y}}$ is suppressed by $\dot{\alpha} / p$ and thus is negligible. Therefore, we give ${\mathcal Y}$ initial conditions
\begin{equation}
{\mathcal Y}_{ij , \, \rm{in}} \simeq \frac{1}{\sqrt{2 \, p_{\rm{in}}}} \, \delta_{ij} \;, \quad \dot{\mathcal Y}_{ij , \, \rm{in}} \simeq - i \sqrt{\frac{p_{\rm{in}}}{2}} \, \delta_{ij} .
\end{equation}

We evolve the system in terms of ${\mathcal Y}$ at the early stage of inflation, i.e. until a few e-folds before $M = p$.%
\footnote{In numerical computation of the early stage of evolution, we needed to Taylor-expand each component of all the matrices up to sufficient order. Without doing this, the components that are relevant for the longitudinal mode of the vector field encounter cancellations in the leading order in the expansion. The sub-leading terms are, however, proportional to $M^2 / p^2$, which is too tiny at early times to be within numerical precision. Such cancellation can be seen in the matrix components $T_{33}$, $X_{3i}$ and $\Omega^2_{3i}$ in Appendix \ref{appA}.}
After this moment, we switch back to the original variables $\delta_i = \left( \delta \varphi , \delta A_x , \delta A_y \right)$, for the ease of computation of the power spectrum. In the next subsection, we briefly formulate this late-time evolution.

\subsection{Late-Time Evolution}

For the late time of evolution (after $M \sim p$), we rotate the system back to the original variables $\delta \varphi$, $\delta A_x$, $\delta A_y$. We do this as it is numerically efficient and the computation of power spectrum is much more compact. We could not do this for the early time, due to a huge hierarchy between the transverse (corresponding to $Y_2$) and longitudinal (to $Y_3$) modes of the vector field. This hierarchy is originated from the fact that $M \ll p$ at the beginning (to have a feeling, if we take $m_0 = 1000 \, m_\varphi$, then $M / p \sim 10^{-76}$ at horizon crossing). The fields $\delta A_x$ and $\delta A_y$ are linear combinations of these modes, and no sensible numerical precision can take such tiny initial values into account. This is why we had to evolve ${\mathcal Y}$ for early times. After $M \gtrsim p$, however, the transverse and longitudinal modes evolve similarly, and thus such an issue no longer arises.

The action for the late time takes the form
\begin{equation}
S_{\rm{2dS}}^{(2)} = \frac{1}{2} \int{dt \, d^3 k} \left( \dot{\delta}^\dagger \, \bar{T} \, \dot{\delta} + \dot{\delta}^\dagger \, \bar{X} \, \delta + \delta^\dagger \, \bar{X}^T \, \dot{\delta} + \delta^\dagger \, \bar{\Omega}^2 \, \delta \right) \label{action_late}
\end{equation}
where $\delta_i = \left( \delta \varphi, \delta A_x, \delta A_y \right)$ and $\bar{T}$ is not a diagonal matrix. This is nothing but a compact expression of (\ref{action_2dS}), and the matrices $\bar{T}$, $\bar{X}$ and $\bar{\Omega}^2$ can easily be identified. As for the early time, the late-time variables must be evolved in the matrix form. We decompose $\delta_i$  by
\begin{equation}
\delta_i \left( t, \vec{k} \right) = \Delta_{ij} \left( t, \vec{k} \right) a_j \left( \vec{k} \right) + \Delta_{ij}^* \left( t, - \vec{k} \right) a_j^\dagger \left( - \vec{k} \right) \label{dec_delta}
\end{equation}
where the creation and annihilation operators are as in (\ref{psipi_h}) and (\ref{dec_Y}). At the transition, we connect the late-time matrices to the early-time ones by
\begin{equation}
\Delta_{ij} = {\mathcal R}_{il} \, {\mathcal Y}_{lj} \; , \quad \dot{\Delta}_{ij} = {\mathcal R}_{il} \, \dot{{\mathcal Y}}_{lj} + \dot{{\mathcal R}}_{il} \, {\mathcal Y}_{lj}.
\end{equation}
The equations of motion for $\Delta_{ij}$ can be found in the manner similar to the early-time case; varying the action (\ref{action_late}), identifying the terms proportional to $a_j \left( \vec{k} \right)$ and $a_j^\dagger \left( - \vec{k} \right)$, we obtain
\begin{equation}
\ddot{\Delta}_{ij} + \bar{{\mathcal A}}_{il} \, \dot{\Delta}_{lj} + \bar{{\mathcal B}}_{il} \, \Delta_{lj} = 0 \label{EOM_Delta}
\end{equation}
where
\begin{equation}
\bar{{\mathcal A}} = \bar{T}^{-1} \left( \dot{\bar{T}} + \bar{X} - \bar{X}^{T} \right) \; , \quad \bar{{\mathcal B}} = \bar{T}^{-1} \left( \bar{\Omega}^2 + \dot{\bar{X}} \right) .
\end{equation}
Using the numerical solution for $\Delta_{ij}$, we compute the power spectrum of primordial curvature perturbation in the vector curvaton scenario. In the next section, we formulate the procedure and show the numerical results of spectrum.

\section{Observables}
\label{sec:observables}

\subsection{Vector Curvaton Scenario}

The mechanism we consider in this paper to produce the curvature perturbations is the curvaton model. It is the mechanism to convert isocurvature (entropy) perturbations into curvature perturbations \cite{Mollerach:1989hu,Lyth:2001nq} (see also \cite{Moroi:2001ct, Linde:1996gt}). In this model, the background expansion is driven by the inflaton ($\phi$, in our case), but the primordial curvature perturbation is generated by some other field ``curvaton'' ($A_\mu$), while the direct contribution from the inflaton perturbation on the final observed cosmological perturbations is negligible, unlike single-field inflation. However, even if the model is arranged such that at late times the contribution of the inflaton perturbations to the cosmological perturbations is much smaller than that of the vector curvaton, the inflaton perturbations can still drastically affect the vector perturbations due to the direct $\varphi-A_\mu$ coupling that necessarily originates from the two functions $f \left( \varphi \right)$ and $m \left( \varphi \right)$.

The physical picture we have in mind is the following: the primordial perturbations of both fields $\varphi$ and $A_\mu$ are originated from quantum fluctuations in the initial adiabatic vacuum. At some time {\emph{during}} inflation, the vector curvaton becomes heavy and starts oscillating; however, due to the coupling with inflaton that breaks the conformal invariance of the vector field, the vector energy density does not decay away even after the oscillation starts but stays nearly constant until the coupling is terminated at the end of inflation. Some time after inflation ends, the inflaton decays into radiation (the detailed mechanism of (p)reheating is beyond the scope of this paper, and we simply assume an instantaneous reheating not long after inflation). The curvaton still oscillates, but it now behaves as non-relativistic dust, since its mass is larger than the Hubble parameter and its coupling to inflaton is no longer present. Since the radiation energy density decrease as $\propto a^{-4}$ and the oscillating curvaton density as $a^{-3}$, the cosmological perturbations induced by those in the radiation density becomes negligible. Due to the coexistence of radiation and dust, the pressure is non-adiabatic; as the curvaton oscillation continues for many Hubble times, this non-adiabatic pressure perturbation converts the curvaton perturbation into the curvature perturbation. Finally (and before  neutrino decoupling \cite{Lyth:2001nq}), the curvaton decays and the curvature perturbation remains constant until the horizon re-enetry. We assume for simplicity that the curvaton decay is also instantaneous.

We consider the curvature perturbation after the universe becomes isotropic (which is true after the vector field becomes heavy), and work in the spatially flat slicing. In this slicing, the curvature perturbation defined on the uniform-density hypersurfaces is given by \cite{Lyth:2001nq, Wands:2000dp}
\begin{equation}
\zeta = - H \frac{\delta \rho}{\dot{\rho}} \simeq \frac{r}{4 + 3 \, r} \frac{\delta \rho_A}{\rho_A} \label{zeta}
\end{equation}
where $\rho = \rho_{\rm r} + \rho_A$ is the total background energy density, $r \equiv \rho_A / \rho_{\rm r}$ is the ratio of the background vector energy density to the radiation density, and $\delta \rho_A$ corresponds to the vector energy density perturbation. In the approximate equality in (\ref{zeta}), we assume that the radiation perturbation has already become negligible at the time of curvaton decay and that $\rho_{\rm r} \propto a^{-4}$ and $\rho_A \propto a^{-3}$. If the curvaton dominates the energy density before its decay, then $\zeta \simeq \frac{1}{3} \frac{\delta \rho_A}{\rho_A}$; in the opposite case, $\zeta \simeq \frac{r}{4} \frac{\delta \rho_A}{\rho_A}$. In either case, the lesson here is that the important quantity we need to focus on is the ratio $\delta \rho_A / \rho_A$.

We can identify the energy density as $- T_0^{0}$, where $T_\mu^{\nu}$ is the energy-momentum tensor of the model, and its perturbation is found by perturbing the field contents. In the limit $\delta g_{\mu\nu} = 0$, we then find
\footnote{The 2D scalar sector is a coupled system and the quantization is done with respect to the initial vacuum, so, in principle, also the cross-correlation  $\left< \delta \rho_A \, \delta \rho_\varphi \right>$ contributes to the value of   $\left< \delta \rho \, \delta \rho \right>$. We verified numerically that the cross correlation is comparable to  $\left< \delta \rho_A \, \delta \rho_A \right>$ at the end of inflation. Since our focus is the study of a curvaton mechanism, we want to assume that $ \delta \rho_\varphi $ plays no role in the post-inflationary perturbation. This can be trivially achieved if, after inflation, the  inflaton  decays into a relativistic species $X$  before the massive vector decays. Then there is a stage in which $\rho_X \propto a^{-4}$, while $\rho_A \propto a^{-3}$, and the cross-correlation decreases  a factor of $a$ faster
 than  $\left< \delta \rho_A \, \delta \rho_\varphi \right>$. }
\begin{equation}
\delta \rho_A = \frac{f}{a^2} \left[ \frac{p_T^2 + M^2}{p^2 + M^2} \,  \dot{A} \, \delta \dot{A}_x 
- \frac{a}{b} \frac{p_L \, p_T}{p^2 + M^2} \, \dot{A} \, \delta \dot{A}_y + M^2 A \, \delta A_x \right] \label{delta_rhoA}
\end{equation}
in the Fourier space. In finding (\ref{delta_rhoA}), $\delta A_0$ is already substituted by (\ref{nondyn}). Notice that (i) the unphysical normalization of scale factors and $A_\mu$ does not affect $\delta \rho$, which is a physical quantity (see the brief discussion after (\ref{constraint})), and (ii) the 2D vector mode $\delta A_z$ does not contribute at linear level  to $\delta \rho$, which is intrinsically a scalar quantity. This justifies disregarding the 2D vector mode.

By using the decomposition (\ref{dec_delta}), we write
\begin{eqnarray}
&& \delta \rho_A \left( t, \vec{k} \right) \equiv \delta \rho_{A,i} \left( t, \vec{k} \right) \, a_i \left( \vec{k} \right) +  \delta \rho_{A,i}^* \left( t, - \vec{k} \right) \, a_i^\dagger \left( - \vec{k} \right) \\
&& \delta \rho_{A,i} =  \frac{f}{a^2} \left[ \frac{p_T^2 + M^2}{p^2 + M^2} \, \dot{A} \, \dot{\Delta}_{2i}
- \frac{a}{b} \frac{p_L \, p_T}{p^2 + M^2} \, \dot{A} \, \dot{\Delta}_{3i} + M^2 \, A \, \Delta_{2i} \right] . 
\end{eqnarray}
The 2-point correlation function of curvature perturbation is $\left< \zeta^2 \right> \propto \left< \delta \rho_A^2 \right> / \rho_A^2$, so we compute
\begin{equation}
\gamma \left( \vec{x} ,\, \vec{y} \right) \equiv
\frac{1}{\rho_A^2 \left( t \right)} \, \left< \delta \rho_{A} \left( t ,\, \vec{x} \right) \,  \delta \rho_{A} \left( t ,\, \vec{y} \right) \right>
=  \int \frac{d^3 k}{\left( 2 \pi \right)^3} \, {\rm e}^{i \vec{k} \cdot \left( \vec{x} - \vec{y} \right)} {\cal F} \left( \vec{k} \right)
\label{corr_int}
\end{equation}
where the expression of $\rho_A$ is shown in (\ref{rho_A}), and
\begin{equation}
{\cal F} \left( \vec{k} \right) \equiv \frac{1}{\rho_A^2} \,  \sum_i \left| \delta \rho_{A,i} \left( \vec{k} \right) \right|^2
\end{equation}
When we chose the momentum orientation in (\ref{k_orient}), we set the coordinate system so that $k_3$ vanish. To be consistent with this choice, we set $\vec{x} - \vec{y} = \left( r_L ,\, r_T ,\, 0 \right)$. Since the function ${\cal F}$ is unchanged under the parity $\vec{k} \rightarrow - \vec{k}$, (\ref{corr_int}) becomes
\begin{equation}
\gamma = \int \frac{d k}{k} \,  \int_0^1 d \xi \, \cos \left( k \, \xi \, r_L \right)
J_0 \left( k \, \sqrt{1-\xi^2} \, r_T \right) \, P_\gamma
\label{corr-aniso}
\end{equation}
where $k = \sqrt{k_L^2 + k_T^2}$, $\xi = k_L / k$, and $J_\nu$ is the Bessel function of the first kind. We have introduced the power spectrum
\begin{equation}
P_\gamma \equiv \frac{k^3}{2 \pi^2} \, {\cal F} \left( k ,\, \xi \right) .
\label{powerspectrum}
\end{equation}
Notice that, in the case of statistical isotropy, ${\cal F}$ depends only on $k$, and we recover the standard expression $\gamma = \int \frac{d k}{k} \, \frac{\sin \left( k \, r \right)}{k \, r} \, P_\gamma$. Finally the quantity of interest is
\begin{equation}
P_\gamma = \frac{k^3}{2 \pi^2} \, \frac{f^2}{a^4 \rho_A^2} \sum_i \left| \frac{p_T^2 + M^2}{p^2 + M^2} \,  \dot{A} \, \dot{\Delta}_{2i} - \frac{a}{b} \, \frac{p_L \, p_T}{p^2 + M^2} \, \dot{A} \, \dot{\Delta}_{3i} + M^2 \, A \, \Delta_{2i} \right|^2 \label{power}
\end{equation}
and in the late-time limit, the middle term in (\ref{power}) is completely negligible and $p \ll M$, so we have
\begin{equation}
P_\gamma \simeq \frac{2 \, k^3}{ \pi^2} \, \sum_i \left| \frac{ \dot{A} \, \dot{\Delta}_{2i} + M^2 \, A \, \Delta_{2i}}{\dot{A}^2 + M^2 \, A^2} \right|^2 \label{power_late}
\end{equation}
where we used (\ref{rho_A}). This is the quantity we compute. In the following subsection, we show the numerical results for $P_\gamma$.

Due to the non-vanishing anisotropy in the background expansion (until the vector field becomes heavy) and due to the scalar-vector interaction, it is natural to consider the possibility that a directional dependence of $P_\gamma$ arises, while it is absent in the simplest single-field inflation. To quantify this statistical anisotropy in the spectrum, we employ the ACW parametrization \cite{Ackerman:2007nb}
\begin{equation}
P \left( \vec{k} \right) = P_{\rm{iso}} \left( k \right) \left[ 1 + g_* \, \xi^2 \right]
\label{P_dec}
\end{equation}
where $\xi$ is the cosine of the angle between the mode $\vec{k}$ and the background privileged direction, coinciding with the $\xi$ above.%
\footnote{
In the case of bispectrum, the parametrization of statistical anisotropy in a vector field model is developed in \cite{Bartolo:2011ee}.
}
This parametrization shares the same concept as the multipole expansion for small higher-order moments. In (\ref{P_dec}), parity ($\vec{k} \rightarrow - \vec{k}$, $\xi \rightarrow - \xi$) symmetry is assumed, which is valid in the model of our interest. Therefore the leading effect of anisotropy is proportional to $\xi^2$, corresponding to a quadrupole. Let us remind that this expansion is valid for small $g_*$, and for $\left| g_* \right| \sim 1$, the contributions from higher-order multipoles become important. The detected  statistical anisotropy that we have mentioned in the Introduction is
$g_* = 0.29 \pm 0.031$ \cite{Hanson:2009gu, Groeneboom:2009cb}, although, as we mentioned, this effect is likely systematic. As a reference, we work under the assumption that the primordial perturbations are statistically isotropic with $\vert g_* \vert < {\rm O } \left( 10^{-1} \right)$. Planck will probe the anisotropy up to $\vert g_* \vert = {\rm O } \left( 10^{-2} \right)$, as shown in \cite{Pullen:2007tu}.

\subsection{Numerical Results}

In this subsection, we show the result of the power spectrum of primordial curvature perturbation (up to overall factor $1/3$ or $r/4$) computed by (\ref{power}) (or its late-time value (\ref{power_late})). As we discussed in Section \ref{sec:background}, the initial equipartition of the vector energy density is required for statistical isotropy in the de Sitter limit with $\delta \varphi = 0$ \cite{Dimopoulos:2009am}, i.e. $\rho_A^{\rm{kin}} = \rho_A^V$, or $\left| C \left( t_{\rm{in}} \right) \right| = \left| D \left( t_{\rm{in}} \right) \right|$. We take this condition for all the numerical computation. For the same reason, the vector mass needs to be $M \ll \dot{\alpha}$ initially and $M \gg \dot{\alpha}$ by the end of inflation, and thus we choose the values of $m_0$ accordingly. Also, to ensure the existence of the attractor (\ref{att_phid}, \ref{att_sigd}, \ref{att_energy}), $c > 1$ is required. We concentrate on the parameter space that satisfies all these conditions. In numerical calculation, we take $\xi = p_L / p$ evaluated at the end of inflation. This coincides with the definition in the previous subsection, since the vector field is already heavy at the end of inflation and the expansion is isotropic from then on.

\begin{figure}
\centering
\includegraphics[width=0.5\textwidth]{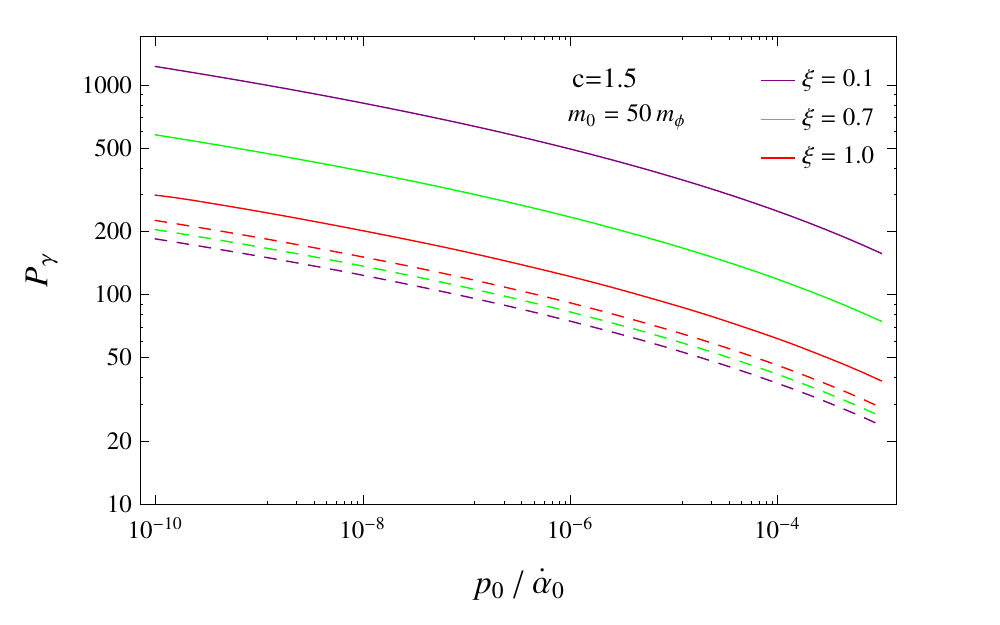}%
\includegraphics[width=0.5\textwidth]{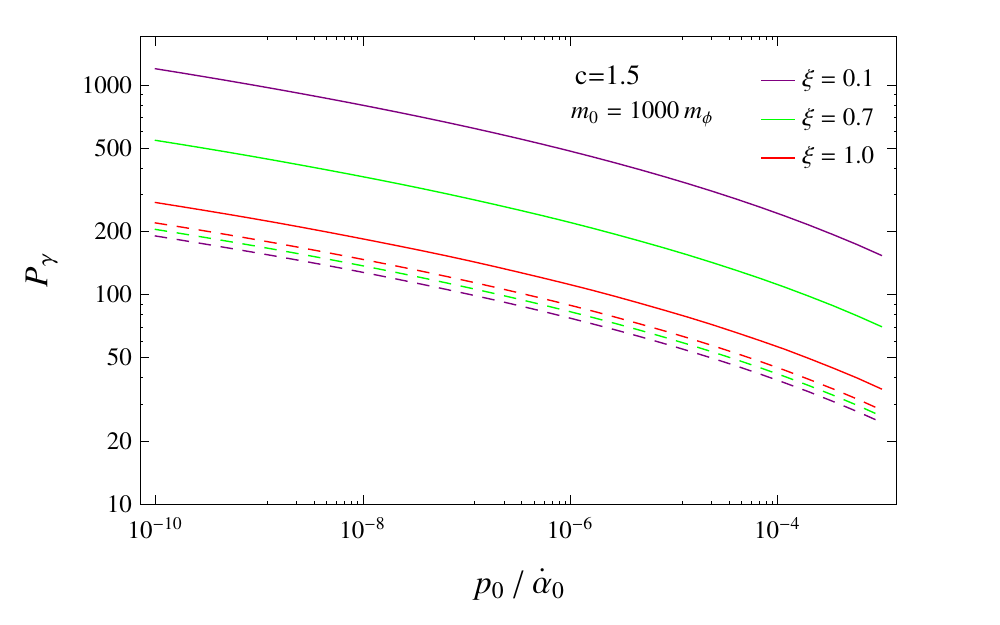}
\caption{Power spectrum $P_\gamma$ defined in (\ref{power_late}), in the units of $m_\varphi^2 / M_p^2$. The left panel shows the result when $m_0 = 50 \, m_\varphi$, and the right when $m_0 = 1000 \, m_\varphi$. In each plot, both the complete results (solid curves) and the cases where the coupling is {\emph{artificially}} neglected (dashed curves) are shown. For each case, a few values of $\xi$ (corresponding to different orientations of $\vec{p}$) are taken to illustrate the directional dependence. Here $c = 1.5$ is arbitrarily chosen to clearly exhibit the difference.}
\label{fig:plot_spec_masscomp}
\end{figure}

\begin{figure}
\centering
\includegraphics[width=0.5\textwidth]{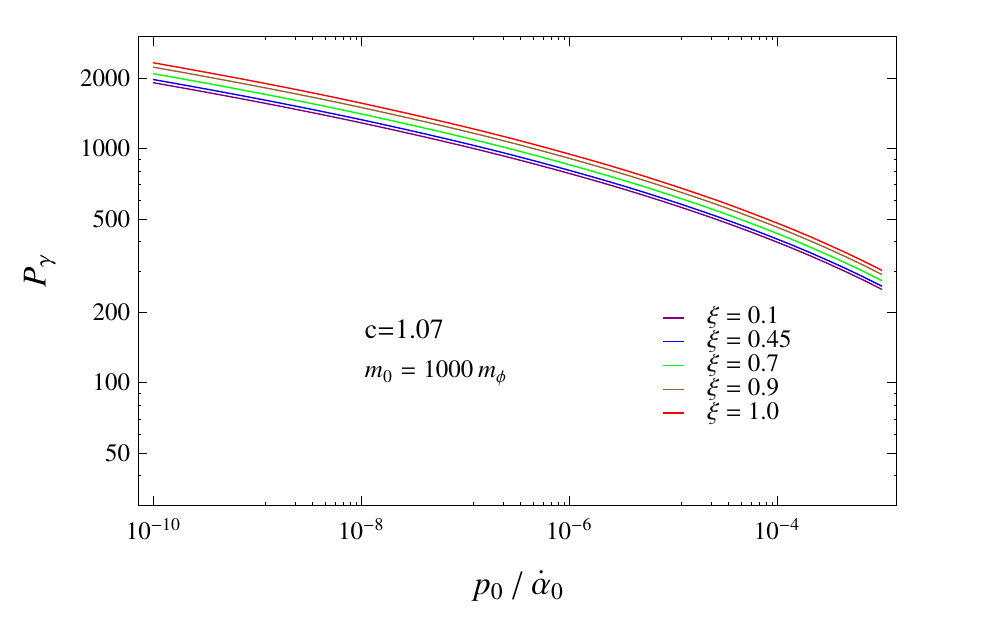}%
\includegraphics[width=0.5\textwidth]{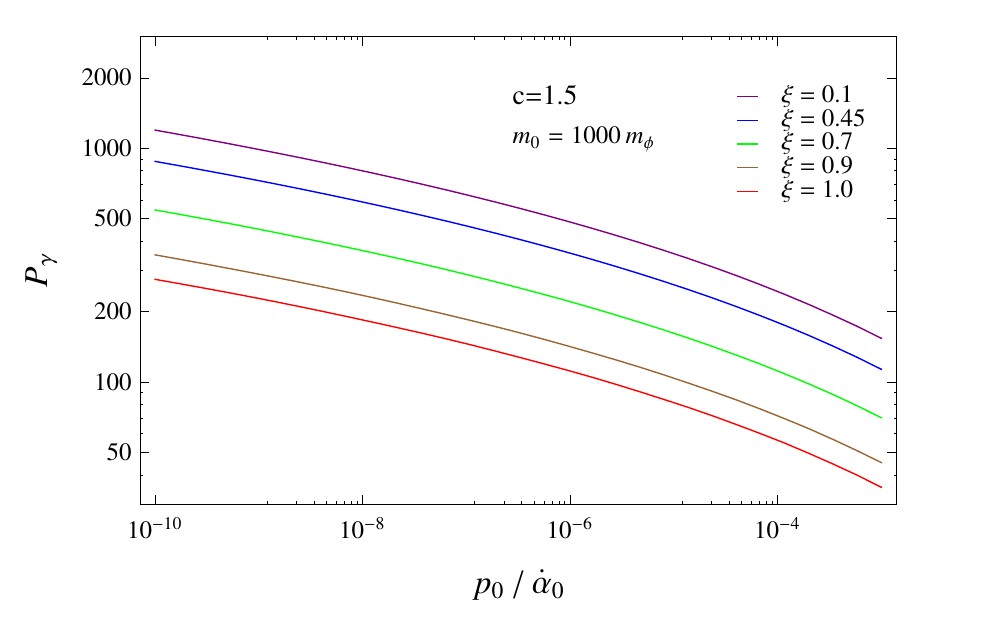}
\caption{Power spectrum $P_\gamma$, in the units of $m_\varphi^2 / M_p^2$. The left panel shows the spectrum when $c = 1.07$, which achieves the observable level of statistical anisotropy, and the right when $c = 1.5$ in order to compare with FIG. \ref{fig:plot_spec_masscomp}. Results for some different values of $\xi$ are shown to visualize the directional dependence and its difference between the two cases. It is observed that the dependence on $\xi$ is flipped between the two panels: larger $\xi$ leads to larger $P_\gamma$ (positive $g_*$) on the left panel, and the opposite (negative $g_*$) on the right. In fact, $g_*$ monotonically decreases with increasing $c$, shown in FIG. \ref{fig:gs_vs_c}. Here the mass of vector field is fixed to be $m_0 = 1000 \, m_\varphi$.}
\label{fig:plot_spec_ccomp}
\end{figure}

Our main interest is the effects due to the non-minimal couplings between the vector and scalar fields introduced in the action (\ref{action}). These couplings are controlled by two parameters $c$ and $m_0$, so in order to see such effects, we observe how the spectrum changes for different values of these parameters. FIGs. \ref{fig:plot_spec_masscomp} and \ref{fig:plot_spec_ccomp} show the computed power spectra for a few different values of $\xi$.%
\footnote{The horizontal axis of FIG. \ref{fig:plot_spec_masscomp} and \ref{fig:plot_spec_ccomp} is $p_0 / \dot{\alpha}_0 \simeq k / \left( a_0 H_0 \right)$, where $0$ denotes the end of inflation. Ideally, we would compute the spectrum of the modes that correspond up to the largest observable scale, or the modes that exit the horizon $60$ e-folds before the end of inflation, but due to numerical difficulties, we show the modes $p_0 / \dot{\alpha}_0 = 10^{-10}$ to $p_0 / \dot{\alpha}_0 = 10^{-3}$ and expect no significant change in the spectral behavior for smaller momenta.}
These choices of $\xi$ are solely of illustrative purpose and have no particular physical significance. In this model, the coupling functions $f$ and $m$ smoothly approach to constants after inflation, but this does not occur instantaneously. As a result, it takes some time for the system to completely decouple and for the curvature perturbation defined in (\ref{zeta}) to become adiabatic direction and freeze out. We therefore evaluate the spectrum a few oscillations (of inflaton) after the end of inflation, when the variation of the values of $P_\gamma$ is at $1\%$ level.

\begin{figure}
\centering
\includegraphics[width=0.55\textwidth]{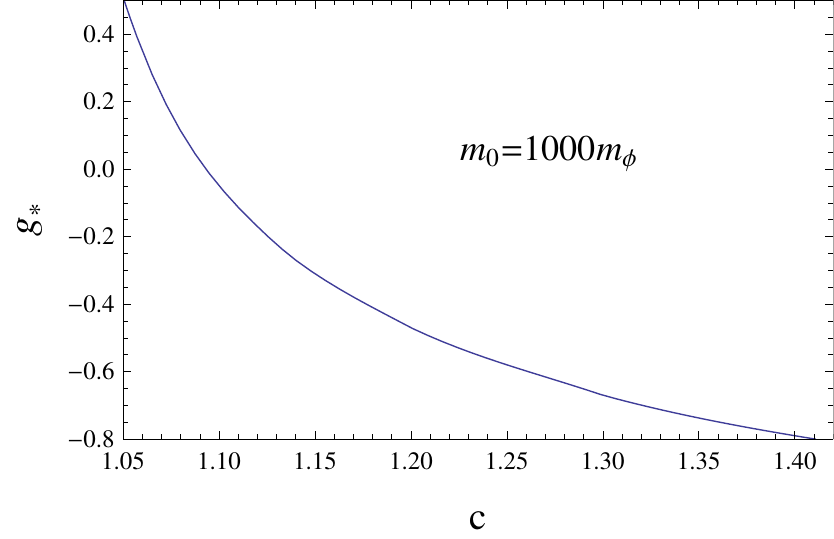}
\caption{The values of $g_*$ as a function of the coupling constant $c$, for $m_0 = 1000 \, m_\varphi$. For smaller values of $m_0$, the curve would shift slightly upward, but the general feature would be unaffected.}
\label{fig:gs_vs_c}
\end{figure}

FIG. \ref{fig:plot_spec_masscomp} shows two cases that compare different values of vector mass: $m_0 = 50 \, m_\varphi$ on the left panel and $m_0 = 1000 \, m_\varphi$ on the right. In each panel, two plots are compared, one (solid curves) with the consistent evolution of the coupled system, and the other (dashed curves) with the vector-scalar coupling neglected for the perturbations (equivalent to setting $\delta \varphi = 0$). Let us emphasize that in the latter, setting $\delta \varphi = 0$ is {\emph{completely artificial}} and is inconsistent within the model. We show this unrealistic case only to illustrate how significant effects the coupling produces, and one can observe that the impact is indeed substantial. If the coupling is manually turned off (dashed curves), the observed or smaller statistical anisotropy can be attained, consistent to the claim in \cite{Dimopoulos:2009am, Dimopoulos:2009vu}
\footnote{
In fact, this is true only for the coupling constant $c \gtrsim 1.25$. If the value of $c$ is closer to $1$, the attractor solutions of background become loose, toward the time when the vector field becomes heavy. More specifically, the equipartition of the vector energy density is not maintained to the sufficient level during the evolution. As is shown in FIG. \ref{fig:gs_vs_c}, the full spectrum with $\delta \varphi \ne 0$ attains unacceptably large anisotropy for $c \gtrsim 1.25$.
}%
; however, if the coupling is present, then the statistical isotropy is clearly violated ($g_* \simeq -0.8$), as can be seen in the figure. Comparing the two panels, we observe that changing the value of the vector mass does not make a significant difference in the spectrum. The overall values of $P_\gamma$ for the coupled case decrease by $< 10 \%$ when $m_0$ is changed from $50 \, m_\varphi$ to $1000 \, m_\varphi$, but the angular dependence does not change: $g_* \simeq -0.8$ for both masses (as a comparison, $g_* \simeq 0.15$ for the $\delta \varphi = 0 $ case). Although this value is undoubtedly excluded by observations, this comparison tells us that (i) the interaction between the vector and scalar through the vector kinetic and mass terms modulates the curvature perturbation significantly, that (ii) satisfying the conditions on the vector mass ($M \ll \dot{\alpha}$ initially and $M \gg \dot{\alpha}$ by the end of inflation) is not sufficient to attain the statistical isotropy, and that (iii) changing the vector mass does not affect the directional dependence of the power spectrum.

The directional dependence of $P_\gamma$ is sensitive to the value of $c$. FIG. \ref{fig:plot_spec_ccomp} compares two cases of different values of $c$ with the same vector mass $m_0 = 1000 \, m_\varphi$. For the purpose of comparison, the right panel takes the same choice of parameters as in the right panel of FIG. \ref{fig:plot_spec_masscomp} (with two more sample values of $\xi$ and without $\delta \varphi = 0$ case). In the left panel, the spectra for the case $c = 1.07$ are shown. For this particular value of $c$, the spectrum acquires the observed level of statistical anisotropy. It is interesting to see that the dependence on $\xi$ is opposite between these two cases. The spectral amplitude is larger for larger $\xi$ in the $c=1.07$ case, and it is smaller for larger $\xi$ in $c=1.5$: this is equivalent to positive $g_*$ for the former and negative for the latter. To quantify $g_*$, we show its values as a function of $c$ in FIG. \ref{fig:gs_vs_c}.%
\footnote{
The angular decomposition of power spectrum, (\ref{P_dec}), is essentially a multipole expansion for small $g_*$. When $\vert g_* \vert$ approaches to $1$, the validity of this expansion starts to break down. The higher-order terms (higher than quadrupole) becomes non-negligible, and indeed, the decomposition (\ref{P_dec}) does not approximate the actual values well. We should treat with care the values of $\vert g_* \vert \sim 1$, but it is clear that such large anisotropy is excluded by the observations.
}
It is found that $g_*$ does not change over the range of momenta.

There are a few things to note. In every case, a nearly scale-invariant power spectrum is obtained, as exhibited in the figures. Secondly, $g_*$ acquires negative values in most of the parameter space, while it becomes positive for the small range $1<c<1.1$. This implies that for $c>1.1$, the larger $p_T$ is, the larger the spectrum, and it is peaked when $\vec{p}$ is completely orthogonal to the direction of the vector vev, i.e. $\xi \rightarrow 0$. Thirdly, the interaction effect is maximal when $\xi$ is close to $0$, as can be seen from FIG. \ref{fig:plot_spec_masscomp}. Since the momentum is completely orthogonal to the  background vector vev at $\xi = 0$, the larger amplitude for $\xi \rightarrow 0$ suggests that the interaction depends largely on $p_T$. Lastly, although the spectrum can achieve the observed anisotropy, the allowed window of $c$ ($g_*$ is insensitive to $m_0$) is very narrow, due to the strong dependence on $c$ near $g_* \simeq 0.3$. This $c$ dependence of the statistical feature of the spectrum is indeed real, resulting from the $\delta \varphi - \delta A_\mu$ interaction.

\section{Discussion and Conclusions}
\label{sec:conclusions}

We studied the primordial inflation with a vector field non-minimally coupled to inflaton through vector kinetic and mass terms, and computed the power spectrum of curvature perturbations within the framework of curvaton mechanism. In this model (\ref{action}), the mass term inevitably introduces a vector longitudinal mode, but it does not acquire a negative kinetic term and therefore is free from ghost instabilities. It is claimed in \cite{Dimopoulos:2009am} that a scale-invariant statistically-isotropic power spectrum can be obtained in this model. However, in their work the anisotropy of background evolution was neglected in the pure de Sitter universe, and the kinetic and mass terms were taken to be external functions of time and thus did not introduce any interaction effect due to the vector-inflaton coupling. In \cite{Dimopoulos:2010xq}, anisotropic background expansion was taken into account, and it was shown that the background attractor leads to the desired time-dependent functions, which we also showed in our language in Section \ref{sec:background}. However, the interaction effects were still assumed to be negligible. In this paper, we relaxed this assumption and did an extensive analysis in the linearized perturbation theory with the consistent inclusion of the vector-inflaton coupling.

The non-trivial features arise due to the presence of the vector vev and its perturbations. The vev of vector field changes the dynamics of the system in three ways: (i) it modifies the background by slowing down the inflaton motion, (ii) it renders the direction of the vev a ``privileged'' direction, that expands differently from the other two, and (iii) it introduces interactions between the inflaton and the vector perturbations already at the linearized level. We have verified that the first two points do not make a dominating change. Namely, if we include (i) and (ii) (not (iii) yet), the anisotropy in the background expansion is small throughout inflation and is rapidly washed away after the vector field becomes heavy; also, the near scale invariance and statistical isotropy of the power spectrum can be attained (see dashed lines in FIG. \ref{fig:plot_spec_masscomp}). The last point, however, generates significant effects on the spectrum. Although the scale invariance is not affected, the statistical isotropy is no longer a generic feature. We stress that this effect was disregarded in all the existing computation of the perturbations in this model.

There are two parameters in the model that control the coupling (\ref{def_fncs}) between the inflaton and vector fields: the inflaton-slowing parameter $c$, and the vector mass end value $m_0$. While changing $m_0$ does not make any appreciable impact on the spectrum, the directional dependence of the spectrum is highly sensitive to the values of $c$. The measure of the statistical anisotropy, $g_*$, is found to be a monotonically decreasing function of $c$ and is mostly negative, except for a small range $1<c<1.1$. This means that for $c>1.1$, the value of spectrum increases as the direction of associated momentum approaches to the direction perpendicular to that of the vector vev. Also, as FIG. \ref{fig:plot_spec_masscomp} suggests, the effect from $\delta A_\mu - \delta \varphi$ coupling is strongest when the momentum is orthogonal to the vector vev. All these features can be justified by observing the action in the early- and late-time regimes. From (\ref{power_late}), we see that only $\delta A_x$ mode contributes (at the linearized level) to the final value of the spectrum, so we look into the relevant part of the action. In the early time, we know $M \ll p$, $\dot{A} \sim M A$, and $\delta A_x \sim \delta A_y \sim a \frac{p}{m} \, \delta \varphi$ (order of magnitude, assuming small background anisotropy). Then the part of the action (\ref{action_2dS}) which contains $\delta A_x$ simplifies to
\begin{eqnarray}
S_{\rm{2dS}}^{(2) \, {\rm{early}}} & \sim & \int{dt d^3 k} \, {\rm e}^{3 \alpha} \frac{f}{2 \, a^2} \Bigg[ \frac{p_T^2}{p^2} \left| \delta \dot{A}_x \right|^2 - \frac{a}{b} \, \frac{p_L \, p_T}{p^2} \left( \delta \dot{A}_x^{\dagger} \, \delta \dot{A}_y + h.c. \right) - p_T^2 \left| \delta A_x \right|^2 + \frac{a}{b} \, p_L p_T \left( \delta A_x^{\dagger} \, \delta A_y + h.c. \right) \nonumber \\
	&& \quad\quad\quad\quad\quad\quad\quad\quad\quad + \frac{f'}{f} \frac{p_T^2}{p^2} \dot{A} \left( \delta \varphi^{\dagger} \, \delta \dot{A}_x + h.c. \right)  \Bigg] + \dots \label{actiondAx_early}
\end{eqnarray}
where $\dots$ denotes the terms without $\delta A_x$. From (\ref{actiondAx_early}), it is seen that initially the interaction term (second line) is negligible, as the initial vacuum is adiabatic, but in the time period when $M \sim p$, this term can make an ${\mathcal O} \left( 1 \right)$ contribution to $\delta A_x$, and thus to the spectrum. This term also illustrates that the interaction strength is proportional to $c$ and $p_T$, and also that it is unrelated to $m_0$.%
\footnote{
One might be concerned that $m_0$ should appear as there is $\dot{A}$ present in the interaction term. However, the value of $\dot{A}$ (and $A$) is enforced by the attractor (\ref{att_energy}). Since $\rho_A = \frac{m^2}{2 a^2} \left( C^2 + D^2 \right)$ and since (\ref{att_energy}) is independent of $m_0$, we have $\left| C \right| = \left| D \right| \propto \frac{1}{m_0}$. Thus the interaction is not affected by the value of $m_0$.
}
In the late time, when the vector mass becomes $M \gg p$, all the directional dependence vanishes from the action, and so no anisotropy is further produced. Therefore, the statistical anisotropy in the spectrum is generated during $M \sim p$ through the interaction between $\delta \varphi$ and $\delta A_x$, and its strength is controlled by $c$ and $p_T$. This heuristic argument justifies the behavior of our numerical results.%
\footnote{
As long as the attractor (\ref{att_energy}) is a good approximation, it is enforced that $\dot{A} \propto \sqrt{c-1} / c$, which implies that the interation term in (\ref{actiondAx_early}) is $\propto \sqrt{c-1}$. Then it appears that as $c \rightarrow 1$, the effects from this term vanish and the statistical isotropy would be achieved. However, in the real evolution, the time variation of $C$ and $D$ (especially the latter) becomes appreciable as $c \rightarrow 1$, and the assumption of constant $C$ and $D$ is violated. Then the simple argument above does not apply; as $c$ gets closer to $1$, $g_*$ becomes positive and large.
}

The vector field is a curvaton and needs to eventually decay to imprint a signature in the curvature perturbations. The curvature power spectrum $P_\zeta$ is either $\frac{1}{3} P_\gamma$ or $\frac{r}{4} P_\gamma$, depending on whether $\rho_A$ or $\rho_{\rm r}$ dominates the total energy density, but the features in the spectrum, such as scale invariance and statistical anisotropy, are directly inherited from $P_\gamma$. While a nearly scale-invariant spectrum is a generic result of this model, statistical isotropy is not. We have shown that the observed value of the anisotropy parameter $g_*$, albeit controversial in itself, can be achieved in this model; however, the allowed parameter space is small (only near $c=1.07$). This is in contrast with the results of \cite{Dimopoulos:2009am, Dimopoulos:2009vu} where statistical isotropy was found to be a result of simply the classical evolution of $f$ and $m$, and with the corresponding claim of \cite{Dimopoulos:2010xq}, in which the results for the perturbations of \cite{Dimopoulos:2009am, Dimopoulos:2009vu} were used, implying statistical isotropy for any value of $c$. This discrepancy is due to the direct coupling between $\delta A_\mu$ and $\delta \varphi$, which was neglected in \cite{Dimopoulos:2009am, Dimopoulos:2009vu}, but should generically be present in any effective field theory realization of the mechanism.

Throughout our analysis, we always assumed the initial equipartition of the vector energy density. In principle, we could compensate the unacceptable statistical anisotropy by adjusting the initial partition. However, since this partition is maintained during the evolution, it would still require fine-tuning. Thus it would not change our conclusion: the statistical isotropy of the curvature power spectrum is not in general attainable without finely-tuned parameters and initial conditions.

Although this model is certainly not the only possibility to realize the vector curvaton scenario with varying kinetic and mass terms of the desired time dependence, we have studied a concrete, realistic example that consistently takes into account the vector-scalar interaction in the presence of the vev of vector field. Curvaton mechanism was first introduced in order to separate the generation of curvature perturbations from the detail of inflation dynamics and to relax the requirements for inflaton \cite{Lyth:2001nq}. This mechanism is favorable in this sense, if the fields are minimally coupled. However, when multiple fields couple to each other in a non-minimal way, as required in  \cite{Dimopoulos:2009am, Dimopoulos:2009vu}, the interaction non-trivially modifies the situation. In such models, calculation in the decoupled limit is not sufficient, and the consistent treatment of the whole system is crucial.

\section*{Acknowledgments}

The author is grateful to  A. Emir G\"{u}mr\"{u}k\c{c}\"{u}o\u{g}lu and Marco Peloso for several useful discussions.  This work of  was supported by the Hoff Lu Fellowship at the University of Minnesota and DOE grant DE-FG02-94ER-40823.

\appendix

\section{Matrices in Early Time Evolution}
\label{appA}

In this appendix, we show the complete expressions of the early-time matrices $T$, $X$ and $\Omega^2$, appearing in (\ref{action_early}). As mentioned in a footnote in the main text, we did not antisymmetrize $X$ for simpler expressions. However, this affects neither the initial conditions at the level of approximation we concern, nor the equations of motion, since antisymmetrization is simply to add total derivatives in the action. On the other hand, $\Omega^2$ is symmetric, by construction.
\begin{equation}
T = \left(\begin{array}{ccc}
1 & 0 & 0 \\
0 & \frac{1}{2} \, \frac{p^2}{p^2 + M^2} \left( 1 + \frac{p_L \, p_T}{\sqrt{p_L^2 + M^2} \sqrt{p_T^2 + M^2}} \right) & 0 \\
0 & 0 & \frac{2 \, p_L^2 p_T^2}{M^2 \left( p^2 + M^2 \right)} \left( 1 - \frac{p_L \, p_T}{\sqrt{p_L^2 + M^2} \sqrt{p_T^2 + M^2}} \right)
\end{array}\right)
\end{equation}
\begin{equation}
X_{11} = - \frac{3}{2} \, \dot{\alpha} \;, \quad X_{12} = 0 \;, \quad X_{13} = 0
\end{equation}
\begin{equation}
X_{21} = \frac{\sqrt{f} \, p}{2  \, a \left( p^2 + M^2 \right)} \dot{A} \frac{f'}{f} \left( \sqrt{p_T^2 + M^2} + \frac{p_L \, p_T}{\sqrt{p_L^2 + M^2}} \right)
\end{equation}
\begin{eqnarray}
X_{22} & = & - \frac{p_L \, p_T + \sqrt{p_L^2 + M^2} \sqrt{p_T^2 + M^2}}{4 \left( p_L^2 + M^2 \right)^{3/2} \left( p_T^2 + M^2 \right)^{3/2} \left( p^2 + M^2 \right)} \Bigg\{ p^2 \left( p_L^2 \, p_T^2 + 2 \, p^2 M^2 + 3 \, M^4 \right) \dot{\alpha} \nonumber \\
	&&  + \left[ 2 \, p_L^2 \, p_T^2 \left( - p_L^2 + 2 \, p_T^2 \right) + \left( - p_L^4 + p_L^2 \, p_T^2 + 2 \, p_T^4 \right) M^2 + 3 \left( - p_L^2 + p_T^2 \right) M^4 \right] \dot{\sigma} \nonumber\\
	&& + p^2 \left[ \left( p_L^2 + M^2 \right) \left( p_T^2 + M^2 \right) \frac{f'}{f} + M' M \left( p^2 + 2 \, M^2 \right) \right] \dot{\phi} \Bigg\}
\end{eqnarray}
\begin{eqnarray}	
X_{23} & = & - \frac{p_L \, p_T \, p \left( p_L \, p_T + \sqrt{p_L^2 + M^2} \sqrt{p_T^2 + M^2} \right)}{2 \left( p_L^2 + M^2 \right)^{3/2} \left( p_T^2 + M^2 \right)^{3/2} \left( p^2 + M^2 \right) } \nonumber \\
	&& \times \left[ \left( p_L^2 - p_T^2 \right) M \, \dot{\alpha} + \left( p_L^2 + 2 \, p_T^2 + 3\, M^2 \right) M \, \dot{\sigma} + \left( p_L^2 - p_T^2 \right) M' \dot{\phi} \right] 
\end{eqnarray}
\begin{equation}
X_{31} = \frac{\sqrt{f} \, p_L \, p_T}{M  \, a \left( p^2 + M^2 \right)} \dot{A} \frac{f'}{f} \left( \sqrt{p_T^2 + M^2} - \frac{p_L \, p_T}{\sqrt{p_L^2 + M^2}} \right)
\end{equation}
\begin{eqnarray}
X_{32} & = & - \frac{p_L \, p_T \, p \left( - p_L \, p_T + \sqrt{p_L^2 + M^2} \sqrt{p_T^2 + M^2} \right)}{2 \left( p_L^2 + M^2 \right)^{3/2} \left( p_T^2 + M^2 \right)^{3/2} \left( p^2 + M^2 \right) } \nonumber \\
	&& \times \left[ \left( p_L^2 - p_T^2 \right) M \, \dot{\alpha} + \left( p_L^2 + 2 \, p_T^2 + 3\, M^2 \right) M \, \dot{\sigma} + \left( p_L^2 - p_T^2 \right) M' \dot{\phi} \right]
\end{eqnarray}
\begin{eqnarray}
X_{33} & = & - \frac{p_L^2 \, p_T^2 \left( - p_L \, p_T + \sqrt{p_L^2 + M^2} \sqrt{p_T^2 + M^2} \right)}{\left( p_L^2 + M^2 \right)^{3/2} \left( p_T^2 + M^2 \right)^{3/2} \left( p^2 + M^2 \right) M^3} \nonumber \\
	&& \times \Bigg\{ \left( 3 \, p_L^2 \, p_T^2 + 4 \, p^2 M^2 + 5 \, M^4 \right) M \, \dot{\alpha} + \left( p_L^2 - 2 \, p_T^2 - M^2 \right) M^3 \, \dot{\sigma} \nonumber \\
	&&  + \left[ \left( p_L^2 + M^2 \right) \left( p_T^2 + M^2 \right) M \frac{f'}{f} + \left( 2 \, p_L^2 \, p_T^2 + 3 \, p^2 \, M^2 + 4 \, M^4 \right) M' \right] \dot{\phi} \Bigg\} 
\end{eqnarray}
\begin{equation}
\Omega^2_{11} = p^2 + V'' - \frac{9}{4} \, \dot{\alpha} + \frac{f}{a^2} \dot{A}^2 \left( \frac{f'^2}{f^2} \, \frac{p_L^2}{p^2 + M^2} - \frac{f''}{2 \, f} \right) + \frac{f}{a^2} M^2 A^2 \left( 2 \frac{f'}{f} \, \frac{M'}{M} + \frac{M'^2}{M^2} + \frac{f''}{2 \, f} + \frac{M''}{M} \right)
\end{equation}
\begin{eqnarray}
\Omega^2_{12} & = & - \frac{\sqrt{f}}{4 \, a \, p \left( p^2 + M^2 \right)} \dot{A} \frac{f'}{f} \nonumber \\
	&& \times \Bigg\{ p_L \, p_T \frac{- p^2 \left( p_L^2 + 3 \, M^2 \right) \dot{\alpha} + 2 p_L^2 \left( p_L^2 - 2 \, p_T^2 + 3 \, M^2 \right) \dot{\sigma} - p^2 \left[ \left( p_L^2 + M^2 \right) \frac{f'}{f} + 2 \, M' M \right] \dot{\phi}}{\left( p_L^2 + M^2 \right)^{3/2}} \nonumber \\
	&& + \frac{- p^2 \left( p_T^2 + 3 \, M^2 \right) \dot{\alpha} + 2 \, p_T^2 \left( p_L^2 - 2 \, p_T^2 - 3 \, M^2 \right) \dot{\sigma} - p^2 \left[ \left( p_T^2 + M^2 \right) \frac{f'}{f} + 2 \, M' M \right] \dot{\phi}}{\sqrt{p_T^2 + M^2}} \Bigg\} \nonumber \\
	&& + \frac{\sqrt{f}}{2 \, a} A \, \frac{p \, M^2}{\sqrt{p_T^2 + M^2}} \left( \frac{f'}{f} + \frac{2 \, M'}{M} \right)
\end{eqnarray}
\begin{eqnarray}
\Omega^2_{13} & = & \frac{\sqrt{f}}{2 \, a} \dot{A} \frac{f^{\prime}}{f} \, \frac{p_L \, p_T}{M^2 \left( p_L^2 + M^2 \right)^{3/2} \sqrt{p_T^2 + M^2} \left( p^2 + M^2 \right)} \nonumber \\
	&& \times \Bigg\{ M \left( p_L^2 + M^2 \right) \left[ \left( M^2 + p_T^2 \right) \sqrt{p_L^2 + M^2} - p_T \, p_L \sqrt{p_T^2 + M^2} \right] \frac{f'}{f} \dot{\phi} \nonumber\\
	&& + 3 \, M \, p_L^2 \, p_T \left( p_T \sqrt{p_L^2 + M^2} - p_L \sqrt{p_T^2 + M^2} \right) \dot{\alpha} 
+ M^5 \sqrt{p_L^2 + M^2} \left( 5 \, \dot{\alpha} + 2 \, \dot{\sigma} \right) \nonumber\\
	&& + M^3 \left[ 3 \, p_T^2 \sqrt{p_L^2 + M^2} \dot{\alpha} + p_L^2 \sqrt{p_L^2 + M^2} \left( 5 \, \dot{\alpha} + 2 \, \dot{\sigma} \right) + p_L \, p_T \sqrt{p_T^2 + M^2} \left( -5 \, \dot{\alpha} + 4 \, \dot{\sigma} \right) \right] \nonumber\\
	&& + 4 \, M^4 M' \sqrt{p_L^2 + M^2} \dot{\phi} + 2 \, p_L^2 \, p_T M' \left( p_T \sqrt{p_L^2 + M^2} - p_L \sqrt{p_T^2 + M^2} \right) \dot{\phi} \nonumber\\
	&& + 2 \, M^2 M' \left[ \left( 2 \, p_L^2 + p_T^2 \right) \sqrt{p_L^2 + M^2} - 2 \, p_L \, p_T \sqrt{p_T^2 + M^2} \right] \dot{\phi} \Bigg\} \nonumber\\ 
	&& + \frac{\sqrt{f}}{a} A \frac{p_L \, p_T \, M}{\sqrt{p_T^2 + M^2}} \left( \frac{f^{\prime}}{f} + \frac{2 \, M^{\prime}}{M} \right)
\end{eqnarray}
\begin{eqnarray}
\Omega^2_{21} & = & \Omega^2_{12} \\
\Omega^2_{22} & = & \frac{p^2}{2} \left( 1 + \frac{p_L \, p_T}{\sqrt{p_L^2 + M^2}\sqrt{p_T^2 + M^2}} \right) - \frac{1}{16 \, p^2 \left( p^2 + M^2 \right)} \nonumber\\
	&& \times \Bigg\{ \left[ \frac{p^2 \left( p_L^2 + 3 \, M^2 \right) \dot{\alpha} - 2 \, p_L^2 \left( p_L^2 - 2 p_T^2 + 3 \, M^2 \right) \dot{\sigma} + p^2 \left( \left( p_L^2 + M^2 \right) \frac{f'}{f} + 2 \, M' M \right) \dot{\phi} }{p_L^2 + M^2}\right]^2 \nonumber\\
	&& + \left[ p^2 \left( p_T^2 + 3 \, M^2 \right) \dot{\alpha} + 2\, p_T^2 \left( 2\, p_T^2 - p_L^2 + 3 \, M^2 \right) \dot{\sigma} + p^2 \left( \left( p_T^2 + M^2 \right) \frac{f'}{f} + 2 M' M \right) \dot{\phi} \right] \nonumber\\
	&& \times \bigg[ 2 \, p_L \, p_T \frac{p^2 \left( p_L^2 + 3\, M^2 \right) \dot{\alpha} - 2 \, p_L^2 \left( p_L^2 - 2 \, p_T^2 + 3\, M^2 \right) \dot{\sigma} + p^2 \left( \left( p_L^2 + M^2 \right) \frac{f'}{f} + 2 M' M \right) \dot{\phi}}{\left( p_L^2 + M^2 \right)^{3/2} \left( p_T^2 + M^2 \right)^{3/2}}  \nonumber\\
	&& + \frac{p^2 \left( p_T^2 + 3 \, M^2 \right) \dot{\alpha} + 2 \, p_T^2 \left( 2 \, p_T^2 - p_L^2 + 3\, M^2 \right) \dot{\sigma} + p^2 \left( \left( p_T^2 + M^2 \right) \frac{f'}{f} + 2 M' M \right) \dot{\phi}}{\left( p_T^2 + M^2 \right)^2 }  \bigg] \Bigg\}
\end{eqnarray}
\begin{eqnarray}
\Omega^2_{23} & = & - \frac{p_L \, p_T \left[ M \left( p_L^2 - p_T^2 \right) \dot{\alpha} + M \left( p_L^2 + 2\, p_T^2 + 3\, M^2 \right) \dot{\sigma} + M' \left( p_L^2 - p_T^2 \right) \dot{\phi} \right]}{2 \, p \left( p^2 + M^2 \right) \left( p_L^2 + M^2 \right)^2 \left( p_T^2 + M^2 \right)^2 M} \nonumber\\
	&& \times \bigg\{ p^2 M \left( p_L^2 + M^2 \right) \left( p_T^2 + M^2 \right) \frac{f'}{f} \dot{\phi} + 3 \, p^4 \, M^3 \, \dot{\alpha} + M^5 \left[ 4 \, p^2 \dot{\alpha} + \left( p_T^2 - 2 \, p_L^2 \right) \dot{\sigma} \right] \nonumber\\
	&& \quad - p_L \, p_T \, M \sqrt{\left( p_L^2 + M^2 \right) \left(p_T^2 + M^2 \right)} \left[ p^2 \, \dot{\alpha} + \left( p_L^2 - 2 \, p_T^2 \right) \dot{\sigma} \right] + p_L^2 \, p_T^2 \, M  \left[ 2 \, p^2 \dot{\alpha} - \left( p_L^2 - 2\, p_T^2 \right) \dot{\sigma} \right] \nonumber\\
	&& \quad + p^2 M' \left[ 3 \, M^4 + 2 \, p^2 \, M^2 + p_L \, p_T \left( p_L \, p_T - \sqrt{\left( p_L^2 + M^2 \right) \left(p_T^2 + M^2 \right)} \right) \right] \dot{\phi}  \bigg\}
\end{eqnarray}
\begin{eqnarray}
\Omega^2_{31} & = & \Omega^2_{13} , \quad \Omega^2_{32} = \Omega^2_{23} \\
\Omega^2_{33} & = & \frac{2 \, p_L^2 \, p_T^2}{M^2} \left[ 1 - \frac{p_L \, p_T}{\sqrt{\left( p_L^2 + M^2 \right) \left(p_T^2 + M^2 \right)}} \right] - \frac{p_L^2 \, p_T^2}{4 \, M^4 \left( p^2 + M^2 \right)} \nonumber\\
	&& \times \Bigg\{ \left[ \frac{ 3 \, p_L^2 \, M \, \dot{\alpha} + M^3 \left( 5 \, \dot{\alpha} - 4 \, \dot{\sigma} \right) + M \left( p_L^2 + M^2 \right) \frac{f'}{f} \dot{\phi} + 2 \, M' \left( p_L^2 + 2 \, M^2 \right) \dot{\phi} }{p_L^2 + M^2} \right]^2 \nonumber\\
	&& - \left[ 3 \, p_L^2 \, M \, \dot{\alpha} + M^3 \left( 5 \, \dot{\alpha} - 4 \, \dot{\sigma} \right) + M \left( p_L^2 + M^2 \right) \frac{f'}{f} \dot{\phi} + 2 \, M' \left( p_L^2 + 2 \, M^2  \right) \dot{\phi} \right] \nonumber\\
	&& \quad \times \left[ M \left( p_T^2 + M^2 \right) \frac{f'}{f} \dot{\phi} + 3 \, p_T^2 M \, \dot{\alpha} + M^3 \left( 5 \, \dot{\alpha} + 2 \, \dot{\sigma} \right) + 2 \, M' \left( p_T^2 + 2 \, M^2 \right) \dot{\phi} \right] \nonumber\\
	&& \quad \times \frac{2 \, p_L \, p_T}{\left( p_L^2 + M^2 \right)^{3/2} \left( p_T^2 + M^2 \right)^{3/2}} \nonumber\\
	&& + \left[ \frac{M \left( p_T^2 + M^2 \right) \frac{f'}{f} \dot{\phi} + 3 \, p_T^2 \, M \dot{\alpha} + M^3 \left( 5 \, \dot{\alpha} + 2 \, \dot{\sigma} \right) + 2 \, M' \left( p_T^2 + 2 \, M^2 \right) \dot{\phi} }{p_T^2 + M^2} \right]^2 \Bigg\}
\end{eqnarray}
It can be seen from the above expressions that all the matrices are real and invariant under the parity $\vec{p} \rightarrow - \vec{p}$.


\end{document}